\documentclass[aps, 
               prb, 
               twocolumn, 
               10pt, 
               nofootinbib,
               amsfonts, 
               amsmath, 
               amssymb]{revtex4-2}

\usepackage{microtype}
\usepackage{upgreek}
\usepackage{graphicx}
\usepackage{hyperref}
\usepackage{xcolor}
\hypersetup{
    colorlinks,
    linkcolor={blue!80!black},
    citecolor={blue!80!black},
    urlcolor={blue!80!black}
}
\usepackage{physics}

\usepackage{placeins}

\usepackage{chemmacros}
\usepackage{siunitx}

\usepackage{xcolor}
\usepackage{ulem}

\begin{document}
\title{Conductance matrix symmetries of multiterminal semiconductor-superconductor devices}

\author{Andrea Maiani}
\affiliation{Center for Quantum Devices, Niels Bohr Institute, University of Copenhagen, DK-2100 Copenhagen, Denmark}

\author{Max Geier}
\affiliation{Center for Quantum Devices, Niels Bohr Institute, University of Copenhagen, DK-2100 Copenhagen, Denmark}

\author{Karsten Flensberg}
\affiliation{Center for Quantum Devices, Niels Bohr Institute, University of Copenhagen, DK-2100 Copenhagen, Denmark}

\date{\today}

\begin{abstract}
Nonlocal tunneling spectroscopy of multiterminal semiconductor-superconductor hybrid devices is a powerful tool to investigate the Andreev bound states below the parent superconducting gap. We examine how to exploit both microscopic and geometrical symmetries of the system to extract information on the normal and Andreev transmission probabilities from the multiterminal electric or thermoelectric differential conductance matrix under the assumption of an electrostatic potential landscape independent of the bias voltages, as well as the absence of leakage currents. These assumptions lead to several symmetry relations on the conductance matrix. Next, by considering a numerical model of a proximitized semiconductor wire with spin-orbit coupling and two normal contacts at its ends, we show how such symmetries can be used to identify the direction and relative strength of Rashba versus Dresselhaus spin-orbit coupling. Finally, we study how a voltage-bias-dependent electrostatic potential as well as quasiparticle leakage breaks the derived symmetry relations and investigate characteristic signatures of these two effects. 
\end{abstract}
\maketitle

\section{Introduction}
Tunneling spectroscopy is a powerful tool for studying superconductor-semiconductor hybrid devices as it provides a clear signature for Andreev bound states (ABS). Nonlocal conductance spectroscopy is the natural extension of local two-probe spectroscopy and overcomes some of its limitations. Initially used in the context of the search for signatures of Cooper-pair splitting~\cite{Hofstetter_Nature_2009, Hofstetter_PRL_2011,  Schindele_PRL_2012, Schindele_PRB_2014, Wang_arXiv_2022}, this type of measurement has been recently considered in the context of topological superconductivity (TS)~\cite{Lobos_NJP_2015, Rosdahl_PRB_2018, Danon_PRL_2020} leading to its use in experiments~\cite{Puglia_PRB_2021, Menard_PRL_2020}, its inclusion in identification protocols for Majorana bound states (MBSs) in nanowires~\cite{Pikulin_arXiv_2021} and unconventional superconductors vortex cores~\cite{Sbierski_PRB_2022}, as well as for the characterization of chiral Majorana edge states in two-dimensional (2D) TS~\cite{Ikegaya_PRL_2019} and the helical gap in two-dimensional electron gases (2DEGs)~\cite{Wojcik_PRB_2021}. Moreover, the same concept appears in experiments involving quantum dots to probe the non-equilibrium dynamics of quasiparticles~\cite{Schindele_PRB_2014, Wang_PRB_2022}. 

When an electron current flows across the device, aside from the electric charge current, energy and heat currents flow too. For this reason, the spectral features of the device, including peaks connected to the onset of the topological phase, can also be identified when analyzing the thermal conductance~\cite{Akhmerov_PRL_2011, Pan_PRB_2021a}. Nevertheless, measurements of thermal transport require a very complex and delicate experimental setup. Easier experiments are the ones that study thermoelectric transport, where the system is driven out of equilibrium by using leads thermalized at different temperatures while the measured output is still a charge current. Thermoelectric measurements have been proposed as an additional tool to investigate subgap features and identify MBSs~\cite{Leijnse_NJP_2014}.

Motivated by recent experimental success in measuring multiterminal electric differential conductance \cite{Hofstetter_PRL_2011, Menard_PRL_2020, Puglia_PRB_2021, Martinez_arXiv_2021, Poschl_arXiv_2022, Wang_arXiv_2022, Banerjee_arXiv_2022a, Banerjee_arXiv_2022b, Banerjee_arXiv_2022c} and the need to characterize hybrid superconductor-semiconductor devices, we here extend the theory of multiterminal tunneling spectroscopy to extract additional information on the electronic and Andreev transmission processes from linear combinations of local and nonlocal differential conductance measurements at different bias voltage or magnetic fields. These linear combinations are derived using conditions that follow from quasiparticle-number conservation, microreversibility, and particle-hole conjugation in the presence of superconductivity. Further relations can be derived in the presence of geometrical symmetries, such as mirror symmetry, or less general Hamiltonian symmetries like additional antiunitary symmetries. 

\begin{figure}[!ht]
    \centering
    \includegraphics[width=\columnwidth]{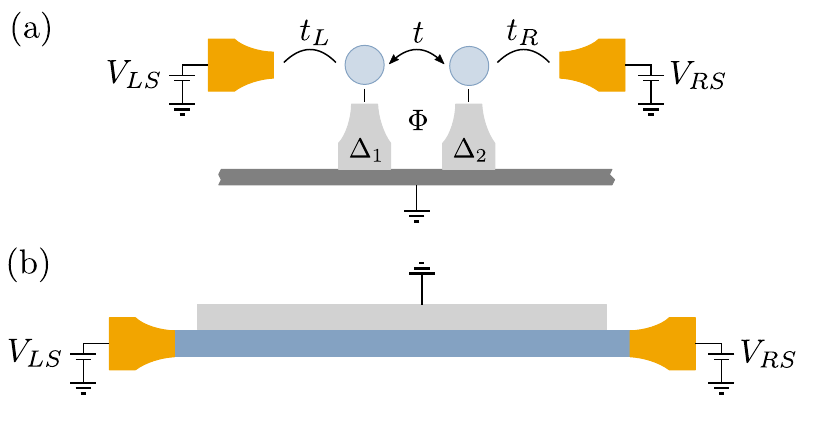}
    \caption{Examples of multiterminal superconductor-semiconductor devices used as toy models in this work: (a) double-dot Josephson junction, and (b) a spin-orbit coupled semiconductor nanowire (blue) proximitized by an $s$-wave superconductor (gray). The metallic contacts are depicted in yellow. All the superconductive leads are grounded.    
    }
    \label{fig:sketches}
\end{figure}

For the specific case of a semiconductor nanowire proximity coupled to an $s$-wave superconductor, we show that the resulting symmetry relations of the conductance matrix can be used to identify the relative strength of Rashba versus Dresselhaus spin-orbit coupling (SOC). We furthermore show that these symmetry relations can be employed to identify signatures of deviations from the assumed symmetries, in particular, voltage bias-dependent electric potential landscapes and quasiparticle dissipation into environmental baths. 

To achieve this objective, we discuss an extended version of the Landauer-Büttiker theory that accounts for bias-voltage-dependent electric potentials. These results are compared to the linear Landauer-Büttiker theory, where it is assumed that the potential landscape, in which the scattering occurs, does not depend on the bias voltages. We refer to this assumption as the \textit{constant landscape approximation} (CLA). 

Extensions of the CLA theory for both nonlinear electric conductance~\cite{Buttiker_JOP_1993, Christen_EPL_1996} and thermoelectric conductance~\cite{Sanchez_PRB_2011, Sanchez_PRL_2013, Yamamoto_PRE_2015} have been considered before. Several previous works focused on the quadratic correction (in voltage bias) to differential conductance obtained by the method of characteristic potential, e.g. \cite{Christen_EPL_1996, Lin_PRB_2001, Meair_JOP_2013, Sanchez_PRL_2013, Hwang_NJP_2013, Texier_PRB_2018}, that in practical applications relies on Thomas-Fermi approximation~\cite{Hernandez_EPJB_2013, Texier_PRB_2018}. This method is more suited for mesoscopic metallic devices with a high density of states, in which the finite-size effects can be neglected. 

In the case of superconductor-semiconductor nanoelectronic devices, instead, the electrostatic potential can be a complicated function of the gate voltage that in general requires the solution of the complete Schr\"odinger-Poisson problem~\cite{Vuik_njp_2016, Dominguez_npjQM_2017, Woods_PRB_2018, Escribano_BJN_2018, Mikkelsen_PRX_2018, Armagnat_scipost_2019, Svejstrup_arXiv_2022}. An initial characterization of finite-bias effects in the fully nonlinear regime was obtained by a combination of approximate analytic and numerical methods in Ref.~\cite{Melo_SciPost_2021}.

This paper is organized as follows. In Sec. \ref{sec:transport_theory}, we describe the general theory of nonlinear charge transport in multiterminal hybrid devices and the difference with CLA results. In Sec. \ref{sec:symmetries}, we show how fundamental symmetries such as microreversibility and particle-hole conjugation in the $S$ matrix generate conductance symmetries valid in CLA and how to use these to extract additional information on the transmission processes from the differential conductance matrix. As an example application, in Sec. \ref{sec:super_semi_soc} we demonstrate how these symmetries can be exploited to identify the spin-orbit direction in a semiconductor nanowire proximitized by a superconductor. Finally, in Sec. \ref{sec:numerics}, with simple numerical simulations we show the effect of finite-bias deformation of the electrostatic potential and dissipation and discuss how violation of conductance symmetry can be used to distinguish between the two. We also illustrate an example of spin-orbit coupling characterization, and discuss how thermoelectric differential conductance can be used as a probe to avoid the finite-bias effect. 

\section{Scattering transport theory}
\label{sec:transport_theory}
The Landauer-Büttiker formalism is a simple yet powerful technique to model transport phenomena. While usually employed in CLA, the nonlinear version can be constructed easily while paying attention to preserving the gauge invariance of the theory~\cite{Christen_EPL_1996}. 

When the motion of quasiparticles is a coherent process and the interactions between quasiparticles beyond mean-field theory can be neglected, quasiparticle transport phenomena in a device are completely described by the single-particle scattering matrix $S$ that relates the probability amplitude of the incoming and outgoing quasiparticles in the leads. More specifically, the relation between the average currents and biases of a multiterminal device depends only on the transmission probability defined as 
\begin{equation}
    T_{\alpha \beta}^{\gamma \delta} (\varepsilon; P) \equiv \tr \abs{S_{\alpha \beta}^{\gamma \delta}(\varepsilon; P) }^2\,,
    \label{eq:transport_def_T}
\end{equation} 
where $S_{\alpha \beta}^{\gamma \delta}$ is the subblock of the scattering matrix that connects the channel of the incoming particles of type $\delta$ from lead $\beta$ to the channel of outgoing particles of type $\gamma$ in lead $\alpha$ for scattering events at energy $\varepsilon$. Since the device we are describing is a superconductor-semiconductor hybrid, the transmission matrix features a Nambu structure with particle and hole sectors, $\delta, \gamma \in \{e, h\}$. The transmission probabilities depend on the set of electric potentials applied to all the electrodes in the system $\{V_\eta\}$ and other external parameters like the applied magnetic field $\vb{B}$ and the pairing amplitude of superconductive leads $\{\Delta_\nu\}$. We denote the set of external parameters by $P=\qty{V_{\eta}}\cup\qty{\Delta_\nu}\cup\qty{\vb{B}}$. In principle also the temperature of the leads can enter as a parameter of the system. For example, temperature could affect the size of the superconductive gap or induce charge accumulation in the semiconductor. We will neglect these effects as we are assuming that the temperature differences involved are much smaller than the critical temperature of the superconductor and are too small to induce a relevant change in the electrostatic potential landscape. Indeed, while voltage bias enters directly into the Poisson equation as boundary conditions, the temperature can enter the electrostatic problem only through charge accumulation.  

A generic multiterminal system comprises a number of normal and superconductive terminals, suggesting a division of the $S$ matrix into subblocks as follows:
\begin{equation}
    S = \begin{pmatrix}
    S_{NN} & S_{NS} \\
    S_{SN} & S_{SS}
    \end{pmatrix}\,.
\end{equation}
In this paper, we will consider the simplest case in which all the superconductive terminals are grounded together and we denote the voltage of the common superconductive lead as $V_S$.

If the system conserves the quasiparticle number, the $S$ matrix is unitary and it follows that 
\begin{align}
    R_{\alpha}^{ee} + R_{\alpha}^{he} + \sum_{\beta\neq\alpha} \qty( T_{\beta \alpha}^{ee} +  T_{\beta \alpha}^{he} ) &= N^e_\alpha(\varepsilon)\,, \label{eq:conservation_in}\\
    R_{\alpha}^{ee} + R_{\alpha}^{eh} + \sum_{\beta\neq\alpha}\qty( T_{\alpha \beta}^{ee} +  T_{\alpha \beta}^{eh} ) &= N^e_\alpha(\varepsilon)\,,  \label{eq:conservation_out}
\end{align}
where we defined for clarity the reflection matrix $R^{\gamma \delta}_{\alpha} \equiv T^{\gamma \delta}_{\alpha \alpha}$ while $N^e_\alpha(\varepsilon)$ is the number of eigenmodes for electrons in lead $\alpha$. Note that, unless the density in the lead is so low to break particle-hole symmetry, $N_\alpha(\varepsilon) \equiv N^e_\alpha(\varepsilon) = N^h_\alpha(\varepsilon)$. Equation~\eqref{eq:conservation_in} represents the conservation of incoming quasiparticles from lead $\alpha$ while Eq.~\eqref{eq:conservation_out} is the same for outgoing quasiparticles. The conservation law breaks down when dissipation effects are included as they do not preserve the particle number. When restricted to energies below the smallest parent superconductor gap $\varepsilon< \min \{\abs{ \Delta_\nu} \}$, the matrices $S_{SS}(\varepsilon)$, $S_{SN}(\varepsilon)$, $S_{NS}(\varepsilon)$ are null and therefore a stronger relation holds where the sum over $\beta$ is restricted to the non-superconductive leads. 

As mentioned above, in the superconductive version of the Landauer-Büttiker theory quasiparticle conservation takes the place of electron conservation for conventional devices. This implies that the electric charge is not explicitly conserved. Indeed, we can distinguish between the charge-conserving normal processes $T^{ee}_{\alpha \beta}$, and $T^{hh}_{\alpha \beta}$, and the non-charge-conserving Andreev processes $T^{eh}_{\alpha \beta}$, and $T^{he}_{\alpha \beta}$. The last two imply, respectively, the creation and the destruction of one Cooper pair in the superconductive leads.

A correct theory of nonlinear conductance of a multiterminal device needs to be gauge invariant. This means that the transmission probabilities are not changed by the addition of a constant offset to all the voltages and particle energy: 
\begin{equation}
     T_{\alpha \beta}^{\gamma \delta} (\varepsilon + \Delta E; \{V_{\eta} + \Delta E\}) 
    = T_{\alpha \beta}^{\gamma \delta} (\varepsilon; \{V_{\eta}\})    
\end{equation}

The easiest way to guarantee gauge invariance is to define a reference voltage. We define the superconductor lead to be our reference voltage and set $V_S = 0$. The first argument of the transmission function is then the energy of the scattering particle with respect to the superconductive lead chemical potential $\varepsilon = E_p + e V_S$. The parameters $V_{\alpha S}$ and $V_{\beta S}$ are the bias of the leads with respect to $V_S$. These biases, together with the voltage biases applied to the capacitively coupled gates, determine the electrostatic potential landscape in which the scattering events take place. Therefore, they enter here as parameters of the $S$ matrix as well as the transmission probability matrix.

\subsection{Currents}
The nonlinear Landauer-Büttiker approach consists in first solving the electrostatic problem for the potential landscape given the applied biases. With the calculated potential landscape, the $S$ matrix for the scattering processes can be evaluated. Finally, once the static scattering problem has been solved, from the $S$ matrix all the transport properties of the system can be derived. In particular, we can use the transmission functions to determine the average currents in the nonequilibrium steady state. In this last step, the terminal biases $V_{\alpha S}$ appear again also as parameters of the distribution functions of the leads.  The application of this approach for superconductive systems has been derived on several occasions~\cite{Blonder_PRB_1982, Takane_1992, Anantram_PRB_1996, Datta_ECETR_1996, Lesovik_PU_2011}, and in the most general formulation the average electric current through a lead $\alpha$ is then 
\begin{equation}
\begin{split}
I^q_{\alpha}(P) = + & \frac{e}{h} \int_{-\infty}^{+\infty}\dd{\varepsilon}\, \qty[f(\varepsilon - e V_{\alpha S},  \theta_{\alpha}) - f(\varepsilon, \theta_{S})] \\ &\times \qty[N_{\alpha} - R_\alpha^{ee}(\varepsilon; P) + R_\alpha^{he}(\varepsilon; P)] \\
- & \sum_{\beta\neq\alpha} \frac{e}{h} \int_{-\infty}^{+\infty}\dd{\varepsilon}\, \qty[f(\varepsilon - e V_{\beta S},  \theta_{\beta}) - f(\varepsilon, \theta_{S})]\\ &\times \qty[T_{\alpha \beta}^{ee}(\varepsilon; P) -  T_{\alpha \beta}^{he}(\varepsilon;P)]\,,
\end{split}
\label{eq:nlqcurrent}
\end{equation}
where $f(\varepsilon, \theta) = \qty(1+e^{\varepsilon/k_B \theta})^{-1}$ is the Fermi-Dirac distribution, and the parameters $\{\theta_\eta\}$ are the temperatures of the leads. We assumed for simplicity that the temperature of all the superconductive leads is equal to $\theta_S$. A similar expression can be written for the energy current $I^\varepsilon_{\alpha}$ while the heat current $I_{\alpha}^{h} = I_{\alpha}^{\varepsilon} + e V_{\alpha S} I_{\alpha}^{q}$ follows easily from the first law of thermodynamics~\cite{Yamamoto_PRE_2015}.

We can write, for the $I^q$ and $I^{h}$ vectors of currents, the following differential relation with the voltages $V$ and temperatures $\theta$ vectors~\cite{Butcher_JOP_1990, Sanchez_PRL_2013, Meair_JOP_2013}:
\begin{equation}
\begin{pmatrix}
\dd I^q_{\alpha} \\
\dd I^h_{\alpha}
\end{pmatrix}
=
\begin{pmatrix}
G_{\alpha \beta} & L_{\alpha \beta} \\
M_{\alpha \beta} & N_{\alpha \beta} 
\end{pmatrix}
\begin{pmatrix}
\dd V_{\beta} \\
\dd \theta_{\beta}
\end{pmatrix}
\end{equation}
where $G$ is the differential electric conductance, $N$ is the differential thermal conductance, while $L$ is the thermoelectric and $M$ is the electrothermal differential conductance. We do not consider thermal transport in this work.

An important observation about the charge current is that it does not satisfy Kirchhoff's current law. This is because the expression derived describes only the quasiparticle current while the supercurrent is not captured in this formalism. By imposing charge conservation, the net supercurrent flowing into the device is $I^{s} = - \sum_\alpha I^q_\alpha$. The net supercurrent can be evaluated as 
\begin{equation}
\begin{split}
I^{s} = - \frac{2e}{h} \sum_{\alpha \beta} \int_{-\infty}^{+\infty}\dd{\varepsilon}\, \qty[f(\varepsilon - e V_{\beta S},  \theta_{\beta}) - f(\varepsilon, \theta_{S})] \\ \times \qty[R_{\beta}^{he}(\varepsilon; P) +  T_{\alpha \beta}^{he}(\varepsilon;P)]\,.
\end{split}
\end{equation}

In the case of a single superconductive lead, this is equal to the supercurrent flowing into the device. In the case of multiple superconductive leads, instead, the supercurrent divides between the different superconductive leads.

\subsection{Constant landscape approximation}
In the CLA, the change in the potential landscape when a voltage bias is applied is neglected. We denote this by writing that $P=P_0$ where $P_0$ is the set of parameters at equilibrium. In this case, the CLA result for electrical conductance is
\begin{equation}
\begin{split}
G_{\alpha \alpha} &= G_0 \int_{-\infty}^{+\infty}\dd{\varepsilon}\, \qty[-\partial_\varepsilon f(\varepsilon - e V_{\alpha S},  \theta_{\alpha})] \\
&\times \qty[N_{\alpha} - R_\alpha^{ee}(\varepsilon; P=P_0) + R_\alpha^{he}(\varepsilon; P=P_0)]\,,
\end{split}
\label{eq:lin_loc_edc}
\end{equation}
for local differential conductance,  while for the nonlocal differential conductance we have
\begin{equation}
\begin{split}
G_{\alpha \beta}& = - G_0 \int_{-\infty}^{+\infty}\dd{\varepsilon}\, \qty[ -\partial_\varepsilon f(\varepsilon - e V_{\beta S},  \theta_{\beta})] \\ & \times \qty[T_{\alpha \beta}^{ee}(\varepsilon; P=P_0) -  T_{\alpha \beta}^{he}(\varepsilon; P=P_0)]\,,
\end{split}
\label{eq:lin_nloc_edc}
\end{equation}
where $G_0=\frac{e^2}{h}$ is the conductance quantum and $\partial_{\varepsilon} f(\varepsilon, \theta) = -\frac{1}{2 k_B \theta} \frac{1}{1 + \cosh (\varepsilon/k_B \theta)}$.

A similar expression can be obtained for thermoelectric conductance. The local and nonlocal thermoelectric conductance reads as
\begin{equation}
\begin{split}
    L_{\alpha \alpha}& =  +  L_0 \int_{-\infty}^{+\infty}\dd{\varepsilon}\, k_B^{-1} \partial_{\theta} f(\varepsilon - e V_{\alpha S},  \theta_\alpha) \\ &\times\qty[N_{\alpha} - R_{\alpha}^{ee}(\varepsilon; P=P_0) + R_{\alpha}^{he}(\varepsilon; P=P_0)]
\end{split}
\label{eq:lin_loc_tedc}
\end{equation}
 and
\begin{equation}
\begin{split}
     L_{\alpha\beta}&=  -  L_0\int_{-\infty}^{+\infty}\dd{\varepsilon}\, k_B^{-1} \partial_{\theta} f(\varepsilon - e V_{\beta S},  \theta_\beta) \\ &\times\qty[T_{\alpha\beta}^{ee}(\varepsilon; P=P_0) -  T_{\alpha\beta}^{he}(\varepsilon; P=P_0)]\,,
\end{split}
\label{eq:lin_nloc_tedc}
\end{equation}
where $L_0 = \frac{e  k_B}{h}$ is the thermoelectric conductance quantum and we used the derivative of the distribution function with respect to temperature, that is $ k_B^{-1} \partial_\theta f(\varepsilon, \theta) = -\frac{\varepsilon}{k_B \theta} \partial_\varepsilon f(\varepsilon, \theta)$. Note that since $\partial_\varepsilon f$ is an odd function of the energy, thermoelectric conductance is sensitive only to the antisymmetric component of the transmission spectrum of the device.

\subsection{Differential conductance in the nonlinear theory}
In principle, the average current given by Eq.~\eqref{eq:nlqcurrent} is exact for a non-interacting system if the potential landscape is calculated self-consistently from the set of parameters $P$. We do not attempt such a calculation here, since the devices treated often have a complicated three-dimensional geometry that cannot easily be reduced to a simple one-dimensional model when taking the electrostatic environment into account. Instead, in this section, we focus on general considerations, while in Sec.~\ref{sec:numerics} we parametrize a potential landscape in a physically motivated way and look at  differences with CLA results.

Evaluating the full derivatives of the charge current $I^q$ with respect to a terminal voltage bias we find that the electric differential conductance can be split into two parts
\begin{equation}
\begin{split}
G_{\alpha \beta}(\qty{V_\eta}) = \dv{I^q_{\alpha}}{V_{\beta S}} = G^\mathrm{(m)}_{\alpha \beta} + G^\mathrm{(def)}_{\alpha \beta}  
\end{split}
\end{equation}
where the first term is the marginal contribution that reads as, i.e. for the local conductance,
\begin{equation}
\begin{split}
G^\mathrm{(m)}_{\alpha \alpha}= + G_0& \int_{-\infty}^{+\infty}\dd{\varepsilon}\, \qty[-\partial_{\varepsilon} f(\varepsilon - e V_{\alpha S}, \theta_{\alpha})] \\
&\times \qty[N_{\alpha} - R_{\alpha }^{ee}(\varepsilon; P) + R_{\alpha}^{he}(\varepsilon; P)] \,.
\end{split}
\end{equation}
This term can be interpreted as the fact that when evaluating the additional current carried by higher-energy states, the $S$ matrix has to be calculated using the potential landscape that takes into account the modified voltage bias. The second term accounts for the deformation of the $S$ matrix for the already filled channels due to the effect of the biasing itself. The deformation contribution reads as
\begin{widetext}
\begin{equation}
\begin{split}
G^\mathrm{(def)}_{\alpha \beta} =  &G_0 \int_{-\infty}^{+\infty}\dd{\varepsilon}\, \qty[f(\varepsilon - e V_{\alpha S}, \theta_\alpha) - f(\varepsilon, \theta_S)] \qty[-\pdv{R_\alpha^{ee}(\varepsilon; P)}{V_{\beta S}} + \pdv{R_\alpha^{he}(\varepsilon; P)}{V_{\beta S}}] \\
+&G_0 \int_{-\infty}^{+\infty}\dd{\varepsilon}\, \qty[f(\varepsilon - e V_{\beta S}, \theta_\beta) - f(\varepsilon, \theta_S)] \qty[-\pdv{T_{\alpha \beta}^{ee}(\varepsilon; P)}{V_{\beta S}} + \pdv{T_{\alpha \beta}^{he}(\varepsilon; P)}{V_{\beta S}}]\,.  
\end{split}
\end{equation}
This correction is often neglected in previous works, e.g. Ref.~\cite{Christen_EPL_1996}. The reason is that in the case of symmetric biasing $V_{\alpha}=V_{\beta}=V$ and fixed temperature $\theta_\alpha=\theta_\beta=\theta_S$, we can rewrite $G^\mathrm{(def)}_{\alpha \beta}$ as 
\begin{equation}
\begin{split}
G^\mathrm{(def)}_{\alpha \beta}(V) = -G_0 \int_{-\infty}^{+\infty}\dd{\varepsilon}\, &\qty[f(\varepsilon - e V, \theta_S) - f(\varepsilon, \theta_S)] \pdv{}{V_\beta} \\
&\times\qty[N_\alpha(\varepsilon) - R_\alpha^{eh}(\varepsilon; P) - R_\alpha^{he}(\varepsilon; P) - \sum_{\beta'} (T_{\alpha \beta'}^{eh}(\varepsilon; P) + T_{\alpha \beta'}^{he}(\varepsilon; P))]\,.  
\end{split}
\end{equation}
Since $N_\alpha$ does not depend on the bias (it is a property of the leads) and all the other terms in the brackets are probabilities of Andreev's processes, it is clear that this quantity vanishes for non-superconductive devices.
\end{widetext}

However, this contribution shows an interesting interplay between electrostatic behavior and superconductivity. 

For example, let us consider a simple N-S junction. The differential conductance can be written as  
\begin{widetext}
\begin{equation}
G_{\alpha \alpha}= + 2 G_0 \int_{-\infty}^{+\infty}\dd{\varepsilon}\, \qty[-\partial_{\varepsilon} f(\varepsilon - e V_{\alpha S}, \theta_{\alpha})] R_{\alpha}^{he}(\varepsilon; P)  +2 G_0\int_{-\infty}^{+\infty}\dd{\varepsilon}\, \qty[f(\varepsilon - e V_{\alpha S}, \theta_\alpha) - f(\varepsilon, \theta_S)] \pdv{R_\alpha^{he}(\varepsilon; P)}{V_{\alpha S}} \,.
\end{equation}
\end{widetext}
It is evident that in case $\pdv{R_\alpha^{he}(\varepsilon; P)}{V_{\alpha S}} > 0$ the second term in the sum can overcome the first one, which is always positive, resulting in negative local differential conductance. A simple case of this can be when the coupling of the scattering region with the superconductor decreases with the bias~\cite{Lesovik_JETP_1998, Melo_SciPost_2021}.

\section{Conductance symmetries}
\label{sec:symmetries}
In this section, we consider how the symmetries of the system manifest themselves first as symmetries of the $S$ matrix and, consequently, in the differential conductance matrix. We consider the ideal CLA case and, for this reason, we drop the biases as arguments in the $S$ matrix.

We choose the Nambu basis of time-reversed holes $\Psi^T = (\psi, \mathcal{T} \psi)^T = \begin{pmatrix} \psi_\uparrow & \psi_\downarrow & -\psi^\dag_\downarrow & \psi^\dag_\uparrow \end{pmatrix}^T$ where we have chosen for the time-reversal symmetry $\mathcal{T} = - i \sigma_y \mathcal{K}$, where $\mathcal{K}$ is the complex-conjugation operator, and therefore for the particle-hole symmetry $\mathcal{P} = i \tau_y \mathcal{T}$.

In the following analysis, we consider symmetric and antisymmetric linear combinations of the conductance matrix elements at opposite voltage bias and magnetic field. We have investigated all linear combinations. However, in the following, we only present the interesting cases in which the linear combination leads to a reduction in the number of terms. 

\subsection{Particle-hole symmetry}
If the system features particle-hole symmetry (PHS), the energy-resolved scattering matrix satisfies the following relation
\begin{equation}
    S(\varepsilon) = \mathcal{P} S(-\varepsilon) \mathcal{P}^\dag =\sigma_y \tau_y S^*(-\varepsilon)  \sigma_y  \tau_y \,.
\end{equation}
This poses an additional constraint on Andreev transmission probabilities that read as
\begin{equation}
T^{\gamma \delta}_{\alpha \beta}(+\varepsilon) = T^{\bar{\gamma} \bar{\delta}}_{\alpha \beta}(-\varepsilon)\,,
\label{eq:phsym}
\end{equation}
where the overbar indicates that the index should be flipped $e \leftrightarrow h$.

This property of the transmission probabilities has a number of consequences on differential conductance. For example, in a system with a single normal terminal (e.g., an NS junction), or for a terminal that is completely isolated from others such that there are no propagating channels (normal or Andreev) connecting it to other leads, the reflection coefficients have to be energy symmetric below the gap. As a consequence, the conductance has to be a symmetric function of the voltage for ideal devices. For the same reason, in these cases the thermoelectric conductance is always exponentially suppressed for temperatures much smaller than the superconducting gap.

Dissipation, inelastic scattering, and coupling to other leads are known effects that break this symmetry of the transmission matrix~\cite{Leadbeater_JOP_1996, Martin_PRB_2014, Liu_PRB_2017} while finite-bias effects can lead to the break-down of the symmetry at conductance-matrix level~\cite{Lesovik_PRB_1997, Melo_SciPost_2021}.

A generalization of this conductance-matrix symmetry for the multiterminal case can be obtained by considering the quantity 
\begin{equation}
    G^\mathrm{sum}_{\alpha}(V) \equiv G_{\alpha\alpha}(V) + \sum_{\beta\neq\alpha} G_{\alpha \beta}(V)\,,
\end{equation} 
that is the sum of the local conductance at terminal $\alpha$ and the nonlocal conductances obtained measuring the current at $\alpha$ while applying a voltage bias to all the other normal leads. It follows that, as a consequence of Eq.~\eqref{eq:phsym},
\begin{equation}
G_\alpha^\mathrm{sum}(V) = G_0 \int_{-\infty}^{+\infty} \dd{\varepsilon} \qty[-\partial_\varepsilon f(\varepsilon - e V)] H_{\alpha}(\varepsilon)\,,
\end{equation} 
where we defined the quantity 
\begin{equation}
\begin{split}
    H_{\alpha}(\varepsilon) = &R_\alpha^{he}(+\varepsilon) + R_\alpha^{he}(-\varepsilon) \\+&\sum_{\beta\neq\alpha} \qty[T_{\alpha \beta}^{he}(+\varepsilon) + T_{\alpha \beta}^{he}(-\varepsilon)]\\
    +&\sum_{\nu } \qty[ T_{\alpha \nu }^{ee}(+\varepsilon)  +T_{\alpha \nu }^{eh}(+\varepsilon) ]\,. 
\end{split}
\end{equation}

The first two terms in $ H_{\alpha}(\varepsilon)$ are explicitly symmetric in $\varepsilon$, while the last sum is null for $\varepsilon<\min{\abs{\Delta_{\nu}}}$. Therefore, as a consequence, 
\begin{equation}
    G^\mathrm{sa}_{\alpha}(V) \equiv G^\mathrm{sum}_{\alpha}(V) - G^\mathrm{sum}_{\alpha}(-V) = 0\,.
    \label{eq:G_a^sa}
\end{equation}
This relation is a generalization of the three-terminal case derived in Ref.~\cite{Danon_PRL_2020}. This result has been derived for non-interacting systems in CLA, therefore, any deviation from zero in $G^\mathrm{sa}$ can be used as a tool to inspect deviations from the CLA and the contributions of quasiparticle dissipation or Coulomb repulsion between quasiparticles. We will discuss these effects in Sec.~\ref{sec:numerics}.

We can split the local and nonlocal differential conductance into symmetric and antisymmetric components
\begin{align}
G^\mathrm{sym}_{\alpha \beta}(V) &\equiv \frac{ G_{\alpha \beta}(V) + G_{\alpha \beta}(- V)}{2}\,, \\
G^\mathrm{anti}_{\alpha \beta}(V) &\equiv \frac{ G_{\alpha \beta}(V) - G_{\alpha \beta}(- V)}{2}\,. 
\end{align}
It has been shown that one can extract the BCS charge, i.e. $\expval{\tau_z}$, of each ABS from the antisymmetric combination $G_{\alpha \beta}^\mathrm{anti}(V)$, given the ABSs are sufficiently separated in the spectrum~\cite{Danon_PRL_2020}.

A similar relation can be derived for thermoelectric differential conductance. Indeed, under the same conditions, we can define 
\begin{equation}
\begin{split}
        L^\mathrm{sum}_{\alpha}(\theta) \equiv  L_{\alpha \alpha} +  \sum_{\beta\neq\alpha} L_{\alpha \beta} \\
        = L_0 \int_{-\infty}^{+\infty} \dd \varepsilon k_B^{-1} \partial_\theta f(\varepsilon, \theta) H_{\alpha}(\varepsilon)\,.
\end{split}
\end{equation}
Since $H_\alpha(\varepsilon)$ is an even function of $\varepsilon$ while $\partial_\theta f(\varepsilon)$ is an odd function, we have that $L^\mathrm{sum}  \simeq 0$ for $k_B \theta\ll\min{\Delta_\nu}$. This holds, again, for non-interacting systems but only for temperatures low enough to exclude excitations of states above the parent gap. Since the thermoelectric conductance is connected to the antisymmetric part of the transmission spectrum, its sign at low temperature can be linked to the BCS charge $\expval{\tau_z}$ of ABSs, following a similar argument as for $G^\mathrm{anti}$ as presented in Ref.~\cite{Danon_PRL_2020}.

In the presence of dissipation, the previous transport symmetries do not hold. Here by dissipation, we mean the presence of a reservoir at the Fermi level that induces quasiparticle leakage. This can be due to various reasons, such as the presence of subgap states in the superconductor causing a softening of the gap or some other leakage mechanism that connects the scattering region to the common ground. A simple way to model quasiparticle leakage is by considering a fictitious lead $\beta'$ that is excluded when taking the calculation of $G^\mathrm{sum}_\alpha$. Focusing on energies below the gap, one finds that the antisymmetric part does not vanish but equals to 
\begin{equation}
\begin{split}
&G^\mathrm{sa}_\alpha(V) =  G_0 \int_{-\infty}^{+\infty} \dd{\varepsilon} \qty[-\partial_\varepsilon f(\varepsilon)] \big[ T_{\alpha \beta'}^{eh}(\varepsilon-eV) \\ + &T_{\alpha \beta'}^{ee}(\varepsilon-eV)
- T_{\alpha \beta'}^{eh}(\varepsilon+eV) - T_{\alpha \beta'}^{ee}(\varepsilon+eV)\big]\,.   
\label{eq:gsa_diss}
\end{split}
\end{equation}
A similar relation can be obtained for thermoelectric conductance. To assess in a more quantitative way the effect of dissipation later we will switch to numerical simulations (see Sec.~\ref{sec:numerics}).

\subsection{Microreversibility}
The microreversibility of the scattering process is a consequence of global time-reversal symmetry and implies that, upon inversion of the time-reversal breaking fields and spin direction, the motion can be reversed\footnote{
Note that we are considering systems that do not break time-reversal symmetry internally.}. As a consequence, the scattering matrix is equal to its transpose
\begin{equation}
    S(\vb B, \Delta_\nu) = \mathcal{T} S(-\vb B, \Delta_\nu^*) \mathcal{T}^\dag = \sigma_y S^T(-\vb B, \Delta_\nu^*) \sigma_y \,,
    \label{eq:sym_S_TRS}
\end{equation}
which, expanding in particle and lead labels, becomes
\begin{equation}
    S^{\gamma \delta}_{\alpha \beta} (\vb B, \Delta_\nu^*) = \sigma_y \qty[S^{\delta \gamma}_{\beta \alpha}(-\vb{B}, \Delta_\nu^*)]^T \sigma_y \,.
\end{equation}

If we consider only non-spin-polarized leads, we can take the trace over the internal spin indices and get the following symmetry relation: 
\begin{equation}
    T^{\gamma \delta}_{\alpha \beta}(\varepsilon, \vb{B}) = T^{\delta \gamma}_{\beta \alpha}(\varepsilon, -\vb{B}, \Delta_\nu^*)\,.
    \label{eq:trsym}
\end{equation}
Combining this relation with the one derived from particle-hole symmetry [Eq.~\eqref{eq:phsym}], we find that, for Andreev reflections and transmission coefficients,
\begin{equation}
    T^{eh}_{\alpha \beta}(\varepsilon, \vb{B}) = T^{eh}_{\beta \alpha}(-\varepsilon, -\vb{B}, \Delta_\nu^*)\,.
    \label{eq:constraint_T_TRS_PHS}
\end{equation}
This has implications both in the electric and thermoelectric nonlocal conductance. 

The connection between microreversibility and thermoelectric quantities has been explored on general grounds both in theory~\cite{Claughton_PRB_1996, Michlek_PRB_2016} and experiments~\cite{Matthews_PRB_2014}. In particular, microreversibility is the microscopic explanation of the Onsager-Casimir relations that lead to other transport symmetries in the charge, heat, and spin channels valid in linear response~\cite{Hartog_PRL_1996, Jacquod_PRB_2012}. 
 
Microreversibilty can be exploited to study separately normal and Andreev processes. To do so, we introduce two new quantities, $G^\mathrm{ra}$ and $G^\mathrm{re}$, both in the local and nonlocal versions, that we call \textit{reciprocal conductances} and that can be extracted from the electric differential conductance matrix:
\begin{align}
G^\mathrm{ra}_{\alpha \beta}(V, \vb{B}) &\equiv G_{\alpha \beta}(V, \vb{B})- G_{\beta \alpha}(+V, -\vb{B}) 
\label{eq:G_ab^ra} \\
G^\mathrm{re}_{\alpha \beta}(V, \vb{B}) &\equiv G_{\alpha \beta}(V, \vb{B})- G_{\beta \alpha}(-V, -\vb{B})
\label{eq:G_ab^re}
\end{align}

\begin{widetext}
By using microreversibility and particle-hole symmetry it is possible to show that
\begin{align}
&G^\mathrm{ra}_{\alpha \beta}(V, \vb{B}) = G_0 \int_{-\infty}^{+\infty} \dd{\varepsilon} \qty[-\partial_\varepsilon f(\varepsilon )]  \qty[ T^{he}_{\alpha\beta} (\varepsilon - e V, \vb{B})- T^{he}_{\alpha\beta} (\varepsilon + e V, \vb{B})]\,, \\
&G^\mathrm{re}_{\alpha \beta}(V, \vb{B}) =  G_0 \int_{-\infty}^{+\infty} \dd{\varepsilon} \qty[-\partial_\varepsilon f(\varepsilon)] \big[ T^{ee}_{\alpha\beta} (\varepsilon - e V, \vb{B}) - T^{ee}_{\alpha\beta} (\varepsilon + eV, \vb{B})\big]\,,
\end{align}
\end{widetext}
where $G^\mathrm{ra}(V)$ is proportional to the antisymmetric part of the Andreev transmission probability while $G^\mathrm{re}(V)$ is proportional to the antisymmetric part of the normal electron transmission probability. For this reason, these two quantities can be used to analyze separately the two types of transport processes. Moreover, it can be verified from their definitions that these two quantities are the decomposition of the antisymmetric part of the local differential conductance:
\begin{equation}
G^\mathrm{anti}_{\alpha \beta} = G^\mathrm{ra}_{\beta \alpha} + G^\mathrm{re}_{\beta \alpha} .
\end{equation}

The local versions, $G^\mathrm{re}_{\alpha \alpha}$ and $G^\mathrm{ra}_{\alpha \alpha}$ are proportional only to the antisymmetric part of the reflection probabilities. As mentioned before, if a lead is sufficiently isolated from the others such that there are no propagating channels connecting it to other leads, the reflection probabilities are bound to be energy-symmetric making the defined quantities null in absence of inelastic scattering.
 
The quantities~\eqref{eq:G_ab^ra}, and~\eqref{eq:G_ab^re} are the only symmetric or antisymmetric combinations of conductance matrix elements $G_{\alpha \beta}(V, \vb{B})$ that simplify to a difference of two transmission or reflection probabilities under the constraints imposed by unitarity [Eqs.~\eqref{eq:conservation_in} and \eqref{eq:conservation_out}], particle-hole symmetry [Eq.~\eqref{eq:phsym}], and time-reversal symmetry [Eq.~\eqref{eq:trsym}]. Contrarily to PHS-derived conductance symmetries, the results in Eqs.~\eqref{eq:G_ab^ra} and \eqref{eq:G_ab^re} are not affected by dissipation since the derivation does not make use of the unitarity of the $S$ matrix.

Note that $\lim_{V\to 0} G^\mathrm{ra}(V) = 0$ in agreement with Onsager-Casimir relation. The vanishing of $G^\mathrm{ra}$ for normal (non-superconductive) devices can be explained as an extension of Onsager-Casimir relations beyond the linear-response regime.

\subsection{Additional antiunitary symmetry}
\label{sec:sym_antiunitary}

Several widely used models in the context of proximitized devices, e.g., the Lutchyn-Oreg Hamiltonian describing a topological phase transition in a proximitized semiconductor nanowire~\cite{Lutchyn_PRL_2010, Oreg_PRL_2010} satisfy an additional antiunitary symmetry $\mathcal{A} = U_\mathcal{A} \mathcal{K}$ aside from microreversibility that persists even in the presence of a Zeeman field. This symmetry implies additional constraints on the conductance matrix. In case the antiunitary symmetry is inherited from the normal state (i.e., it holds separately for electron and hole parts of the wavefunction), then the matrix $U_\mathcal{A}$ does not mix the particle-hole and lead indices. In this case, the symmetry condition for the scattering matrix can be written as
\begin{equation}
 S(\vb{B},\Delta_\nu)=U_\mathcal{A}^T S(\vb{B},\Delta_\nu)^T U_\mathcal{A}^*  \,.
\end{equation}
As a consequence, the transmission probabilities satisfy the symmetry relations
\begin{equation}
    T^{\gamma \delta}_{\alpha \beta}(\vb{B},\Delta_\nu) = T^{\delta \gamma}_{\beta \alpha}(\vb{B},\Delta_\nu)\,,
\end{equation}
The validity of this symmetry on the transmission probabilities is due to the block-diagonal structure of the unitary $U_\mathcal{A}$ combined with the definition of $T^{\gamma \delta}_{\alpha \beta}(\vb{B}, \Delta_\nu)$ in Eq.~\eqref{eq:transport_def_T} that contains a trace over all single-lead indices that are present in the normal state. 

In combination with PHS [Eq.~\eqref{eq:phsym}], we find 
\begin{equation}
     T^{\gamma \delta}_{\alpha \beta}(+\varepsilon, +\vb{B},\Delta_\nu) = T^{\bar{\delta} \bar{\gamma}}_{\beta \alpha}(-\varepsilon, +\vb{B},\Delta_\nu)\,.   
\end{equation}
and, in particular for the Andreev transmission,
\begin{equation}
    T^{eh}_{\alpha \beta}(+\varepsilon, +\vb{B},\Delta_\nu) = T^{eh}_{\beta \alpha}(-\varepsilon, +\vb{B},\Delta_\nu)\,. 
\end{equation}  

The combination with microreversibility in Eq.~\eqref{eq:trsym} instead gives
\begin{equation}
     T^{\gamma \delta}_{\alpha \beta}(+\varepsilon, +\vb{B},\Delta_\nu) = T^{\gamma \delta}_{\alpha \beta}(+\varepsilon, -\vb{B},\Delta_\nu^*)\,.   
\end{equation}

As a result, the \textit{conductance magnetic asymmetry}, that we define as
\begin{equation}
      G^\mathrm{m}_{\alpha \beta}(V, \vb{B},\Delta_\nu) \equiv G_{\alpha \beta}(V, \vb{B},\Delta_\nu) - G_{\alpha \beta}(V, -\vb{B},\Delta_\nu^*)\,.
    \label{eq:G^m_ab}
\end{equation}
vanishes.\footnote{Note that for local quantities $G^\mathrm{m}_{\alpha \alpha} = G^\mathrm{ra}_{\alpha \alpha}$.} Violations of this symmetry relation can be attributed to perturbations that break the antiunitary symmetry $\mathcal{A}$. This can be the result o, e.g., orbital effects or phase inhomogeneities. 

Similar considerations can be drawn for  thermoelectric conductance. In the same fashion, we can define the \textit{thermoelectric conductance magnetic asymmetry} 
\begin{equation}
    L^\mathrm{m}_{\alpha \beta}(V, \vb{B},\Delta_\nu) \equiv L_{\alpha \beta}(V, \vb{B},\Delta_\nu) - L_{\alpha \beta}(V, -\vb{B},\Delta_\nu^*)\,,    
\end{equation}
this quantity vanishes under the same assumptions.

Note that, in simpler systems like two-terminal metallic wires, the conductance is expected to be a symmetric function of the magnetic field. Therefore, the presence of a conductance magnetic asymmetry can be used as a probe of electron-electron interactions in the system~\cite{Texier_PRB_2018}. In multiterminal superconductive devices instead, the symmetry is not expected even in the non-interacting system where the presence of the additional antiunitary symmetry is necessary to have a vanishing $G^\mathrm{m}$.

\subsection{Geometrical symmetries}
Geometrical symmetries of the device can also be exploited to build quantities that select only specific components of the transmission matrix. These can be useful in case the geometry of the system can be controlled to some degree of accuracy such that it may feature approximate geometrical symmetries.

For introducing a concrete example, let us consider a 2DEG wire like the one in Figs.~\ref{fig:sketches}(a) and \ref{fig:supersem_soc_sketches}(a). The wire is aligned along the $x$ direction, with two symmetric leads $L$ and $R$. Suppose the system is symmetric upon mirroring along the $x$ and $y$ directions, $\mathcal{M}_x$, and $\mathcal{M}_y$, and it features a rotation symmetry $\mathcal{R}_z(\pi)$. By using only $\mathcal{T}$, $\mathcal{P}$ and $\mathcal{M}_x$ symmetries we have 
\begin{equation}
    T^{eh}_{RL} (\varepsilon, B_x, B_y, B_z) \xrightarrow{\mathcal{T}\mathcal{P}\mathcal{M}_x} T^{eh}_{RL} (-\varepsilon, -B_x, B_y, B_z) \\
    \label{eq:geom_sym_1}
\end{equation}
\begin{equation}
    T^{ee}_{RL} (\varepsilon, B_x, B_y, B_z) 
    \xrightarrow{\mathcal{T}\mathcal{M}_x} T^{ee}_{RL} (\varepsilon, -B_x, B_y, B_z) 
    \label{eq:geom_sym_2}
\end{equation}
and we can use this result to define new quantities similar to the reciprocal differential conductances, but with the advantage that is evaluated at only one lead and reversing only one component of the magnetic field
\begin{align}
G^\mathrm{xa}_{LR}(V, \vb{B})&\equiv G_{LR}(V, B_x) - G_{LR}(V, -B_x)\,,  \\  
G^\mathrm{xe}_{LR}(V, \vb{B}) &\equiv G_{LR}(V, B_x) - G_{LR}(-V, -B_x)\,. 
\end{align}

\begin{widetext}
By using Eqs.~\eqref{eq:geom_sym_1} and \eqref{eq:geom_sym_2} it is possible to show that
\begin{align}   &G^\mathrm{xa}_{LR}(V, \vb{B}) = G_0 \int_{-\infty}^{+\infty} \dd{\varepsilon} \qty[-\partial_\varepsilon f(\varepsilon )]  \qty[ T^{he}_{LR} (\varepsilon - e V, \vb{B})- T^{he}_{LR} (\varepsilon + e V, \vb{B})] \\
&G^\mathrm{xe}_{LR}(V, \vb{B}) =  G_0 \int_{-\infty}^{+\infty} \dd{\varepsilon} \qty[-\partial_\varepsilon f(\varepsilon)] \big[ T^{ee}_{LR} (\varepsilon + e V, \vb{B}) - T^{ee}_{LR} (\varepsilon - eV, \vb{B})\big],
\end{align}
\end{widetext}

Similarly, for the thermoelectric conductance, we have
\begin{equation}
L^\mathrm{x}_{LR}(\theta, B_x) \equiv L_{LR}(\theta, B_x) - L_{LR}(\theta, - B_x)\simeq 0\,.
\label{eq:Lr}
\end{equation}
Again, this quantity is exactly zero if we neglect the energies above the parent gap and thus deviation at low temperatures can be directly linked to dissipation effects.

If a system featuring mirror symmetry satisfies an additional antiunitary symmetry as discussed in Sec.~\ref{sec:sym_antiunitary}, then it follows that $G^\mathrm{ra}_{\alpha \beta} = 0$. In absence of the additional antiunitary symmetry, the conductance symmetry $G^\mathrm{ra}_{\alpha \beta} = 0$ is present in case the magnetic field lies in the plane orthogonal to the mirror symmetry axis. These properties can be used as an indication of whether the system satisfies mirror symmetry.

\section{Additional antiunitary symmetry in a proximitized semiconductor nanowire}
\label{sec:super_semi_soc}
\begin{figure}[!ht]
    \centering
    \includegraphics[width=\columnwidth]{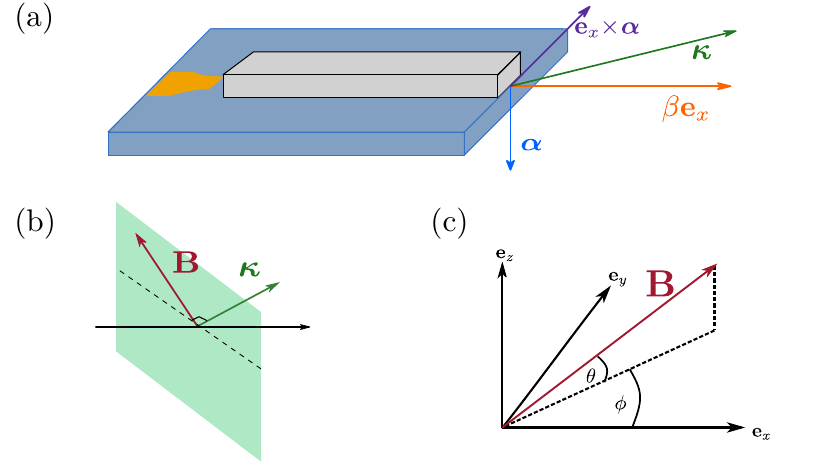}
    \caption{Sketch of a proximitized semiconductor nanowire with spin-orbit coupling. 
    In (a), using as a reference a simple 2DEG geometry, we introduce a Rashba field $\vb*{\alpha}$ transverse to the nanowire direction $\vb{e}_\mathrm{x}$ due to the interface electric field. The Rashba SOC manifest in a momentum-dependent spin splitting in the $\vb{e}_\mathrm{x} \times \vb*{\alpha}$ direction. A native Dresselhaus spin-orbit coupling can also be present; this acts as momentum-dependent $\vb{e}_\mathrm{x}$ spin splitting. The combination of the two effects can be described by a generalized spin-orbit direction $\vb*{\kappa}$ that defines an orthogonal plane shown in green in (b). When the magnetic field $\vb{B}$ lays in the orthogonal plane, the system features an additional antiunitary symmetry. In (c) we introduce a coordinate system to discuss the dependence of transport on the direction of the external magnetic field $\vb{B}$.
    }    
    \label{fig:supersem_soc_sketches}
\end{figure}
To introduce a concrete example application of the additional antiunitary symmetry discussed in Sec.~\ref{sec:sym_antiunitary}, we now consider a quasi-1D semiconductor nanowire proximitized by a superconductor. An important question for these devices is the characterization of spin-orbit coupling. This can be achieved by leveraging the symmetry relations previously introduced.

We consider a system represented by the following low-energy effective Hamiltonian
\begin{equation}
\begin{split}
    \mathcal{H}= &\qty[\frac{\hbar^2 \vb{k}^2}{2 m^*}  + V(\vb{r}) + \mathcal{H}_\mathrm{SOC}] \tau_z \\
    &+\vb{b} \cdot \vb*{\sigma} \tau_0 + \tilde{\Delta} \tau_x \,,  
\end{split}
\label{eq:supersemi_ham}
\end{equation}
where $\vb{k} = - i \grad$ is the wavevector, $m^*$ is the effective mass, $V(x)=-\mu(x)$ is the potential landscape (which can include disorder), $\tilde{\Delta}$ is the proximity-induced pairing potential in the weak coupling limit, and $\vb{b}$ is the Zeeman spin splitting in the semiconductor. Finally, the spin-orbit coupling $\mathcal{H}_\mathrm{SOC} = \mathcal{H}_R + \mathcal{H}_D$ is the sum of the Rashba and the Dresselhaus term. 

The Rashba interaction reads $\mathcal{H}_{R} = \vb{k} \cross \vb*{\alpha} \cdot \vb*{\sigma}$ where the Rashba field $\vb*{\alpha}$ is proportional to the electric field in the device. The Dresselhaus spin-orbit coupling arises from the lack of inversion symmetry of the material and can be written as $\mathcal{H}_{D} = \gamma_D \vb{l} \cdot \vb*{\sigma}$. In zinc-blende crystals, the $\vb{l}$ vector components are $l_a = k_a (k_b^2 - k_c^2)$ where $(a,b,c)$ are cyclic permutations of the coordinates $(x,y,z)$~\cite{Campos_PRB_2018}.

Here we consider a quasi-1D system in the $\vb{e}_x$ direction as shown in Fig.~\ref{fig:supersem_soc_sketches}. For sufficiently thin wires with sufficiently energy-separated eigenmodes with different radial momenta, we can replace the radial momentum operators by their expectation values evaluated on the transverse eigenmode wave function $\vb{k} \simeq (k_x, \expval{k_y}, \expval{k_z})$. Under this assumption, we can rewrite the Dresselhaus Hamiltonian as $\mathcal{H}_{D} = \gamma_D \sigma_x k_x  \qty[- \langle k_z^2 \rangle + \langle k_y^2 \rangle] = \beta k_x \sigma_x$, while for the Rashba SOC, $\mathcal{H}_R = (\alpha_{y} \sigma_z - \alpha_{z} \sigma_y) k_x$, where a term $ \langle \vb{k}_\perp \rangle \cross \vb*{\alpha}_\parallel \cdot \vb*{\sigma}$ vanishes due to $\langle \vb{k}_\perp \rangle = 0$ for confined eigenmodes. 

First, consider the case of pure Rashba spin-orbit coupling, i.e., $\gamma_D = 0$. The Hamiltonian satisfies an antiunitary symmetry when the magnetic field points within the plane spanned by the Rashba field $\vb*{\alpha}$ and the direction of the wire. For $\vb*{\alpha} = \alpha \vb{e}_z$, the antiunitary symmetry is complex conjugation and the real-space Hamiltonian in Eq.~\eqref{eq:supersemi_ham} is real.

If the system features both Rashba and Dresselhaus spin-orbit coupling, the plane spanned by the magnetic fields that preserve an antiunitary symmetry is tilted. Without loss of generality we choose a Rashba field perpendicular to the wire pointing along $z$, i.e. $\vb*{\alpha}_\perp = \alpha_\perp \hat{e}_z$.  We introduce a coordinate system for the magnetic field defined as $\vb{b} = b \qty(\cos \theta \cos \phi, \cos\theta \sin\phi, \sin\theta)$ where $\theta$ is the elevation and $\phi$ the azimuth with respect to the wire direction. In this case, the spin-orbit coupling term reads $k_x (\alpha_\perp \sigma_y + \beta \sigma_x)$. A rotation $e^{i \phi_\kappa \sigma_z/2}$ in spin space by the angle $\tan \phi_\kappa= \frac{\beta}{\alpha_\perp}$ transforms $e^{- i \phi_\kappa\sigma_z/2} k_x (\alpha_\perp \sigma_y + \beta \sigma_x) e^{i \phi_\kappa\sigma_z/2} = k_x \sqrt{\alpha_\perp^2 + \beta^2} \sigma_y$. In this basis, spin-orbit coupling is real and the Hamiltonian satisfies $\mathcal{A} = \mathcal{K}$. 
      
This antiunitary symmetry is preserved by a Zeeman field $\vb{b} = b_\perp \sigma_z + b_\parallel (\cos \phi_\kappa\sigma_x - \sin\phi_\kappa\sigma_y)$, such that $e^{- i \phi_\kappa\sigma_z/2} \left[b_\perp \sigma_z + b_\parallel (\cos \phi_\kappa\sigma_x - \sin \phi_\kappa\sigma_y) \right] e^{i \phi_\kappa\sigma_z/2} = b_\perp \sigma_z + b_\parallel \sigma_x$ is real. Here, $\vb{b}_\perp$ is the component of the magnetic field parallel to $\vb*{\alpha}_\perp$ and $\vb{b}_\parallel$ is the component pointing in the direction orthogonal to $\vb*{\alpha}_\perp$ and $\beta \hat{x} + \alpha_\perp \hat{y}$. In other words, the combined Rashba and Dresselhaus spin-orbit coupling terms can be written as $\vb*{\kappa} \cdot \vb*{\sigma} k_x$ where $\vb*{\kappa} = \vb{e}_{x} \times \vb*{\alpha}_\perp + \beta \vb{e}_{x}$ is the generalized spin-orbit direction. The antiunitary symmetry is preserved by a Zeeman field $\vb{b}\cdot\vb*{\kappa} = 0$. The plane spanned by the $\vb{b}$ vectors that satisfy the orthogonality condition can be identified by an angle $\phi_0 = \phi_\kappa + \pi/2$.

\section{Numerical models}
\label{sec:numerics}
In this section, we introduce two numerical models to show examples of how microscopic symmetries of the systems manifest themselves in the transport properties. In both cases, we model the grounded leads in the system with the method of self-energies. This can be useful when the lead is a metal with high density compared to the scattering region as the self-energy takes the simple form of a local complex-valued potential. This is added to the Hamiltonian to generate an effective non-Hermitian energy-dependent Hamiltonian that can be studied with the scattering approach.

We consider the most general case of a grounded soft-gap superconductor that can be described by the Dynes superconductor model~\cite{Herman_PRB_2016, Herman_PRB_2018, Kavicky_PRB_2020}. In the case of a superconductive lead with a high density of states, the intermediate coupling regime can be adequately described by the following local self-energy 
\begin{equation}
\Sigma_\nu(\varepsilon, \vb r) = \gamma_\nu(\vb r) \frac{- \qty(\varepsilon + i \Gamma_\nu) \tau_0\sigma_0 + \qty(\Delta_\nu \tau_+ + \Delta_\nu^\dag \tau_-)\sigma_0}{\sqrt{\abs{\Delta_\nu}^2\tau_0\sigma_0 - \qty[(\varepsilon  + i \Gamma_\nu)\tau_0 \sigma_0]^2}}\,,    
\label{eq:numerics_self_energy}
\end{equation}
where $\Delta_\nu = |\Delta_\nu| e^{i \phi_\nu}$ is the pairing amplitude with phase $\phi_\nu$ (that we assume constant in space), $\Gamma_\nu$ is the Dynes parameter that models pair-breaking scattering processes, the local coupling strength is $\gamma_\nu = \pi \mathcal{D}_{\nu} t_\nu^2$ with $\mathcal{D}_{\nu}$ being the density of states in the lead $\nu$, and $t_\nu$ is the interface hopping amplitude between the scattering region and the superconductive lead $\nu$. For a normal lead, this reduces to
\begin{equation}
    \Sigma_\mathrm{n}(\varepsilon, \vb r) = - i \gamma_\mathrm{n}(\vb{r}) \sigma_0 \tau_0 \,,
    \label{eq:numerics_self_energy_n}
\end{equation}
that is an imaginary potential that causes the decay of the quasiparticle wave function.

\subsection{Double-dot Josephson junction}
To study the effect of the additional antiunitary symmetry discussed in Sec.~\ref{sec:sym_antiunitary}, we first consider a double-dot Josephson junction illustrated in Fig.~\ref{fig:sketches}. The effective Hamiltonian of the system is 
\begin{equation}
\begin{split}
    \mathcal{H} = &\begin{pmatrix}
    -\mu_1 & -t \\
    -t & -\mu_2 
    \end{pmatrix} \tau_z 
    \\
    &+\begin{pmatrix}
    1 & 0 \\
    0 & 0
    \end{pmatrix} \Sigma_1(\varepsilon) 
    +
    \begin{pmatrix}
    0 & 0 \\
    0 & 1
    \end{pmatrix} \Sigma_2(\varepsilon)
\end{split}
\end{equation}
where $\mu_i$ are the local chemical potentials in the dots, $t$ is the hopping amplitude between the dots, and $\Sigma_i$ are the local self-energies induced by the superconductive leads $i = 1,2$. The scattering matrix can be obtained by using the Weidenmüller formula (see, e.g., \cite{Dittes_PR_2000})
\begin{equation}
    S(\varepsilon) = \mathbb{I} - 2\pi i W^\dag \frac{1}{\varepsilon-\mathcal{H}
    +\mathrm{i}\pi W W^\dag} W\,,
\end{equation}
where $W_i (E) \equiv \sqrt{\rho_i (\varepsilon)} t_i \Pi_i(\varepsilon)$ with $t_i$ the tunneling amplitude from the device to lead $i$, $\rho_i(\varepsilon)$ the density of states in lead $i$, and $\Pi_i(\varepsilon)$ the projector onto the eigenstates of lead $i$ at energy $\varepsilon$. In our model, we approximate the tunnel coupling between lead $L$ and dot $1$ (lead $R$ and dot $2$) by two energy-independent parameters, such that $W = \begin{pmatrix} w_L & w_R \end{pmatrix}\tau_z $.

\begin{figure*}[!ht]
    \centering
    \includegraphics[width=\textwidth]{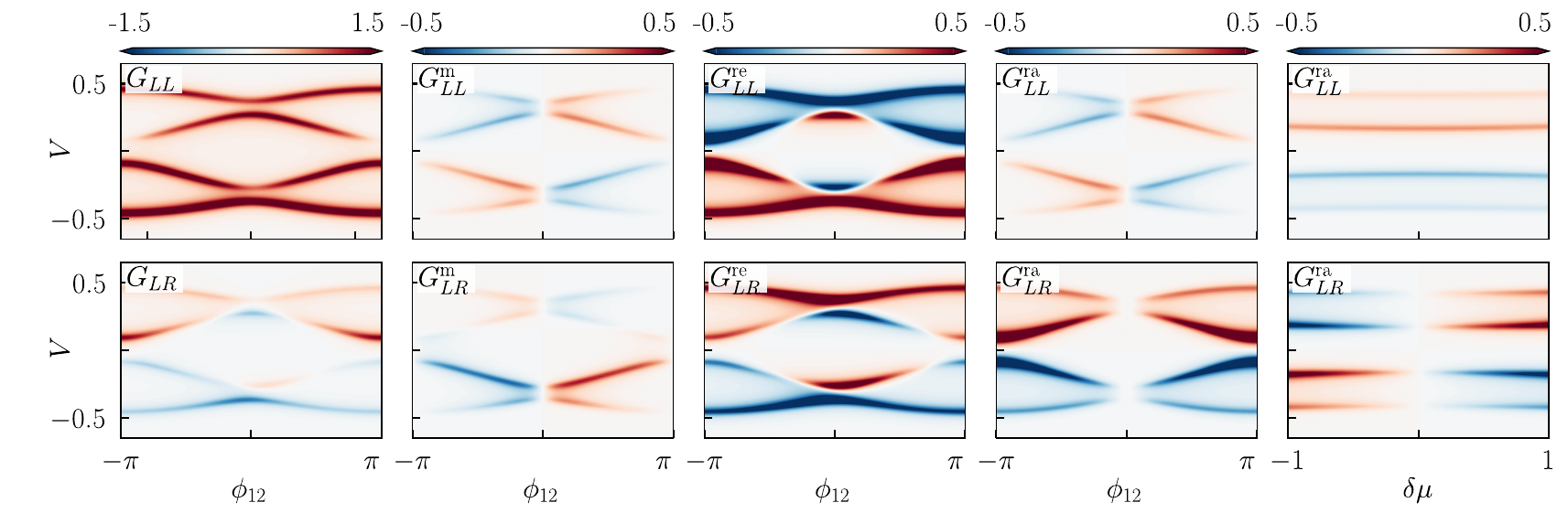}
    \caption{
    Local conductance $G_{LL}$ and combinations $G_{LL}^\mathrm{m}$, $G_{LL}^\mathrm{re}$, $G_{LL}^\mathrm{ra}$ (top row, from left to right) and non-local conductance $G_{LR}$ and combinations $G_{LR}^\mathrm{m}$, $G_{LR}^\mathrm{re}$, $G_{LR}^\mathrm{ra}$ (bottom row, from left to right) in a double-dot Josephson junction as a function of the phase difference $\phi_{12} = \phi_1 - \phi_2$ for a symmetric configuration $\mu_1 = \mu_2$ (columns one to four). 
    The rightmost column displays the reciprocal conductance $G_{LL}^{ra}$ for $\phi_{LR} = \pi/4$ as a function of the chemical potential asymmetry $\delta\mu = (\mu_1 - \mu_2) / (\mu_1 + \mu_2)$. 
    The parameters used are $t=\SI{0.2}{\milli\eV}$, $w_L^2=w_R^2=\SI{0.01}{\milli\eV}$, $\mu_1=\mu (1 + \delta\mu)$, $\mu_2=\mu (1 - \delta\mu)$,
    $\mu=\SI{0.10}{\milli\eV}$, $\gamma_1=\gamma_2=\SI{0.3}{\milli\eV}$, $|\Delta_1| = |\Delta_2| = \SI{1}{\milli\eV}$. 
    }
    \label{fig:double dot}
\end{figure*}

When the phase difference between the two superconductive terminals $\phi_{12} = \phi_{1} - \phi_{2}$ is zero or $\pi$, the system features the antiunitary symmetry $\mathcal{A} = \mathcal{K}$. As a consequence, the Andreev process probabilities are symmetric in the energy axis, i.e., $T^{eh}(\varepsilon) = T^{eh}(-\varepsilon)$, and $R^{eh}(\varepsilon)=R^{eh}(-\varepsilon)$. This can be verified by the zero in the conductance magnetic asymmetry $G^\mathrm{m}$, as shown in the second column of Fig.~\ref{fig:double dot}. 

The reciprocal conductance can also be used to verify the presence of a mirror symmetry of the device. A mirror symmetry $\mathcal{M}_x$ exchanges the two dots and reverses the sign of the phase difference $\phi_i \to - \phi_i$. The latter can be seen by noticing that the phase difference can be created by a magnetic field $B_z$ piercing a superconducting loop within the $x$-$y$-plane connecting to the two dots. This mirror symmetry implies a zero in the non-local reciprocal conductances $G^\mathrm{ra}_{LR}$, $G^\mathrm{ra}_{RL}$. Indeed, by calculating $G^\mathrm{ra}_{LR}$ as a function of the dots' levels asymmetry $\delta\mu = (\mu_1 - \mu_2) / (\mu_1 + \mu_2)$ (see the rightmost column of Fig.~\ref{fig:double dot}), we verify the presence of a zero in $G^\mathrm{ra}$ for $\delta\mu=0$, that is when the system feature a mirror symmetry. 

As a function of phase difference $\phi_{12}$, the symmetric configuration $\mu_1 = \mu_2$ exhibits a zero in the reciprocal conductances $G^\mathrm{ra}_{LL}$, $G^\mathrm{ra}_{LR}$ at $\phi_{12} = 0$ due to the mirror symmetry (c.f. the fourth column of Fig.~\ref{fig:double dot}). The zero in $G^\mathrm{ra}_{LL}$, $G^\mathrm{ra}_{LR}$ signals energy-symmetric Andreev reflection and transmission amplitudes [c.f. Eq. \ref{eq:G_ab^ra}]. This mirror symmetry is broken at finite $|\phi_{12}| > 0$, including $\phi_{12} = \pi$, due to the different superconducting phase at the two dots. 

\subsection{Proximitized semiconductor nanowire}
\label{sec:num_wire}

\begin{figure}[!hb]
    \centering
    \includegraphics[width=\columnwidth]{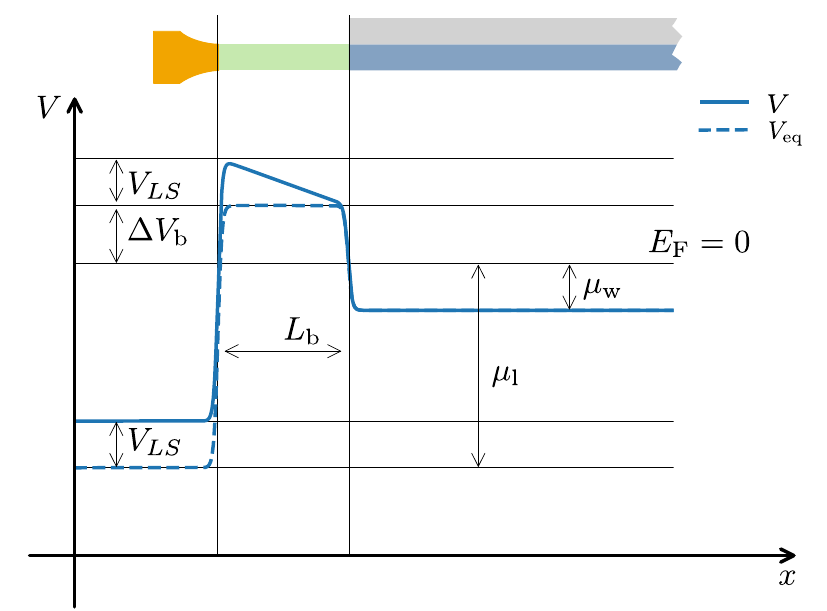}
    \caption{Sketch of one contact of a three-terminals device. The monomodal wire (blue, green) is coated with a superconductive shell (gray), which induces a superconducting pairing potential in the central region and fixes the chemical potential due to the high density. Two barriers (quantum point contacts) are introduced on the side (green), while barrier gates can partially control the shape of the potential drop here. Two bias voltages are applied on the left and right leads (yellow) with respect to the grounded superconductor. In the lower panel, we sketch the potential landscape of the system $V(x)$ in the unbiased regime (dashed blue line) and biased one (solid blue line)  showing all the quantities used to parametrize the potential landscape. 
    }
    \label{fig:nanowire_sketch}
\end{figure}

As an example of a three-terminal device, we consider the case of a semiconductor nanowire proximitized by an $s$-wave superconductor as shown in Fig. \ref{fig:nanowire_sketch}. We demonstrate how an antiunitary symmetry persisting at a finite magnetic field for specific directions can be employed to extract the ratio between Dresselhaus and Rashba spin-orbit coupling, as introduced in Sec.~\ref{sec:super_semi_soc}. We further study the effects of dissipation and voltage-bias-dependent potentials on the symmetry relations derived under CLA.

The Hamiltonian is similar to the one in Eq. \eqref{eq:supersemi_ham}, but here we treat the superconductive lead using the self-energy model
\begin{equation}
\begin{split}
\mathcal{H}(\varepsilon) = &\qty[\frac{\hbar^2 k_x^2}{2 m^*}  + V(x)] \tau_z \\
    &+\qty[(\alpha_{y} \sigma_z - \alpha_{z} \sigma_y) k_x + \beta k_x \sigma_x] \tau_z\\
    &+\vb{b} \cdot \vb*{\sigma} \tau_0 + \Sigma(\varepsilon) \,, 
\label{eq:nanowire_ham_numeric}
\end{split}
\end{equation}
where we take $m^* = 0.026 m_e$, consistent with an InAs nanowire, and $\Sigma(\varepsilon)$ is the superconductive lead self-energy as given in Eq. \eqref{eq:numerics_self_energy}.

To include the effect of a finite bias, we model the deformation of the potential landscape in a simplified effective manner. Following an approach proposed by Ref.~\cite{Melo_SciPost_2021}, we assume that the density of states in the superconductor shell is high enough to guarantee perfect screening of the electric field. This means that the potential drop falls entirely in the (depleted) barrier region. The case of imperfect screening is analyzed in Ref.~\cite{Pikulin_arXiv_2021}.

To provide a gauge-invariant description of the potential landscape we select the voltage applied to the superconductor as the reference voltage $V_\mathrm{S} = 0$ such that $\varepsilon = E_{p} - \mu_S = E_p + e V_\mathrm{S}$ is the energy of the scattering particle. We define the left and right biases as $V_{LS}$ and $V_{RS}$. The other parameters that enter in the effective potential landscape are the chemical potential in the lead and the wire, that we define as $\mu_\mathrm{l} = -e(V_\mathrm{l} - V_\mathrm{S})$ and $\mu_\mathrm{w} = -e(V_\mathrm{w} - V_\mathrm{S})$, and the zero-bias barrier height $\Delta V_\mathrm{b} = e(V_\mathrm{b} - V_\mathrm{S})$. We assume the absence of built-in biases in the junction by considering a flat potential barrier at zero bias. The effect of the zero-bias barrier $\Delta V_\mathrm{b}$ can be connected to a reduction of the coupling with the leads that causes a reduction of the height of the peaks and increased sharpness in the differential conductance. All these parameters are shown in the sketch of the landscape shown in Fig.~\ref{fig:nanowire_sketch}. We modeled the effect of the finite bias as a linear voltage drop (i.e., a constant electric field) within the barrier, and we smoothed the potential using a sigmoid function instead of Heaviside steps to avoid sharp transitions between the different parts of the system.  

We used a nanowire length of $L_\mathrm{w}=\SI{500}{\nano\m}$ with barriers of length $L_\mathrm{b} = \SI{50}{\nano\m}$. The local chemical potential in the nanowire is set to $\mu_\mathrm{w} = \SI{0.5}{\milli\eV}$ while the zero-bias barrier height is set to $\Delta V_\mathrm{b} = \SI{0.3}{\milli\eV}$. The lead have $\mu_\mathrm{l} = \SI{25}{\milli\eV}$. For the superconductive lead, we set $\Delta = \SI{0.35}{\milli\eV}$ and $\gamma_{\rm Sc} = \SI{0.2}{\milli\eV}$.
To simplify the evaluation of reciprocal conductance, we restrict the elevation angle $\theta \in [-\pi/2, \pi/2]$ while allowing the magnitude $b$ to take negative values.
We discretized the Hamiltonian using the finite-differences method with step lengths $a_x = \SI{1}{\nano\meter}$, then evaluated the scattering matrix $S[\varepsilon, \mathcal{H}(\varepsilon; P)]$ using the \texttt{Kwant} package for quantum transport~\cite{Groth_NJP_2014}. After evaluation of the $S$ matrix, the conductance is calculated for the CLA case following Eqs.~\eqref{eq:lin_loc_edc}-\eqref{eq:lin_nloc_tedc}, while in the nonlinear case, the electric charge current is calculated by numerical integration of Eq.~\eqref{eq:nlqcurrent}. 

\subsubsection{Identification of the spin-orbit coupling direction}
\label{sec:num_wire_SOC}

We first focus on the newly introduced quantities, reciprocal differential conductances and conductance magnetic asymmetry, and their use for the determination of the spin-orbit coupling direction. To emphasize the effect and maximize $G^\mathrm{ra}$, we consider a strongly asymmetric case in which the left barrier is set to $\Delta V_{\mathrm{b}, L}=\SI{0.3}{\milli\eV}$ while the right barrier is in the open regime $\Delta V_{\mathrm{b}, R}=0$. We also choose to align the Rashba field in the out-of-plane direction $\alpha_R = (0, 0, -10)\,\si{\milli\eV \nano\meter}$ while we set the Dresselhaus energy to $\beta=\SI{5}{\milli\eV \nano\meter}$. 

\begin{figure}[!ht]
    \centering
    \includegraphics{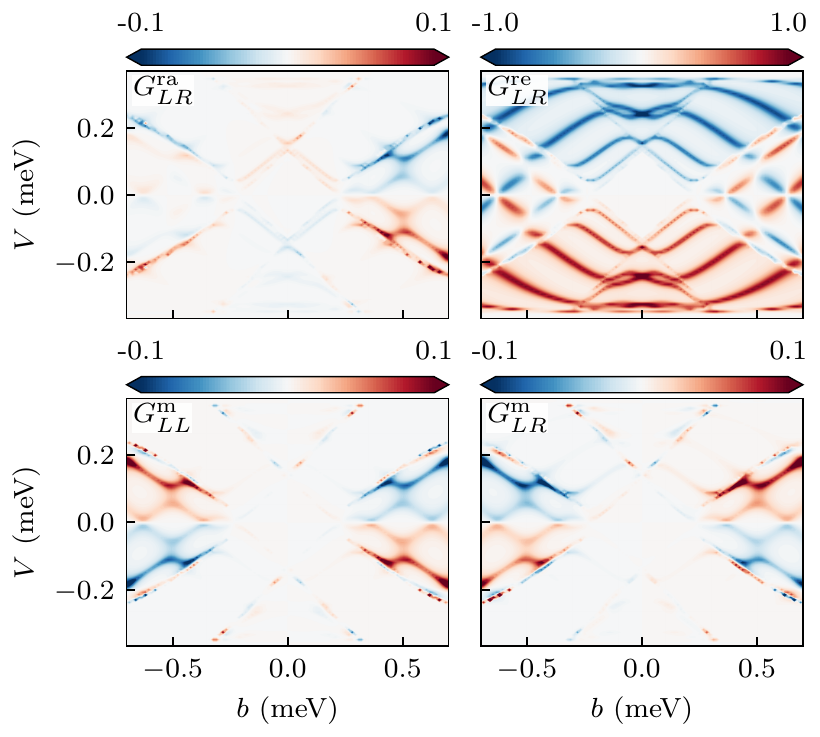}
    \caption{Reciprocal conductance and conductance magnetic asymmetry in a super-semi nanowire.
    We show the typical signature of the reciprocal conductance and conductance magnetic asymmetry [Eqs.~\eqref{eq:G_ab^ra},~\eqref{eq:G_ab^re}, and~\eqref{eq:G^m_ab}] for a proximitized semiconductor nanowire with Rashba spin-orbit coupling as a function of the Zeeman energy $b$. We set the direction of the magnetic field as $\theta=\frac{\pi}{4}$ and $\phi=0$.
    In this simulation, $V_{\mathrm{b},L}=\SI{30}{\micro\eV}$, $V_{\mathrm{b},R}=\SI{0}{\micro\eV}$, and $\beta=\SI{5}{\milli\eV \nano\meter}$.}
    \label{fig:reciprocal}
\end{figure}

A sweep in Zeeman energy $b$ is shown in Fig.~\ref{fig:reciprocal}. We find that $G^\mathrm{re}$ is much larger than $G^\mathrm{ra}$. By Eqs.~\eqref{eq:G_ab^ra} and \eqref{eq:G_ab^re}, this indicates that the antisymmetric part of the electron-electron transmission probability dominates over the antisymmetric part of the crossed Andreev reflection probability. Note that in the presence of both mirror symmetry $\mathcal{M}_x$ (inverting the wire direction) and an antiunitary symmetry, the Andreev transmission probabilities are symmetric in energy such that $G^\mathrm{ra}_{LR}$ vanishes. For our device, in the presence of only Rashba SOC and a magnetic field oriented in the wire direction, a mirror-symmetric device satisfies an antiunitary symmetry $\mathcal{A} = \mathcal{K}$ and a mirror symmetry $\mathcal{M}_x = \sigma_x \tau_0$. Therefore, a signal in $G^\mathrm{ra}_{LR}$ is correlated to the mirror symmetry breaking terms in the device geometry, such as the asymmetric barrier configuration used here, and Dresselhaus spin-orbit coupling $\propto k_x \sigma_x$, breaking both the antiunitary symmetry and the mirror symmetry.

\begin{figure}[!hb]
    \centering
    \includegraphics{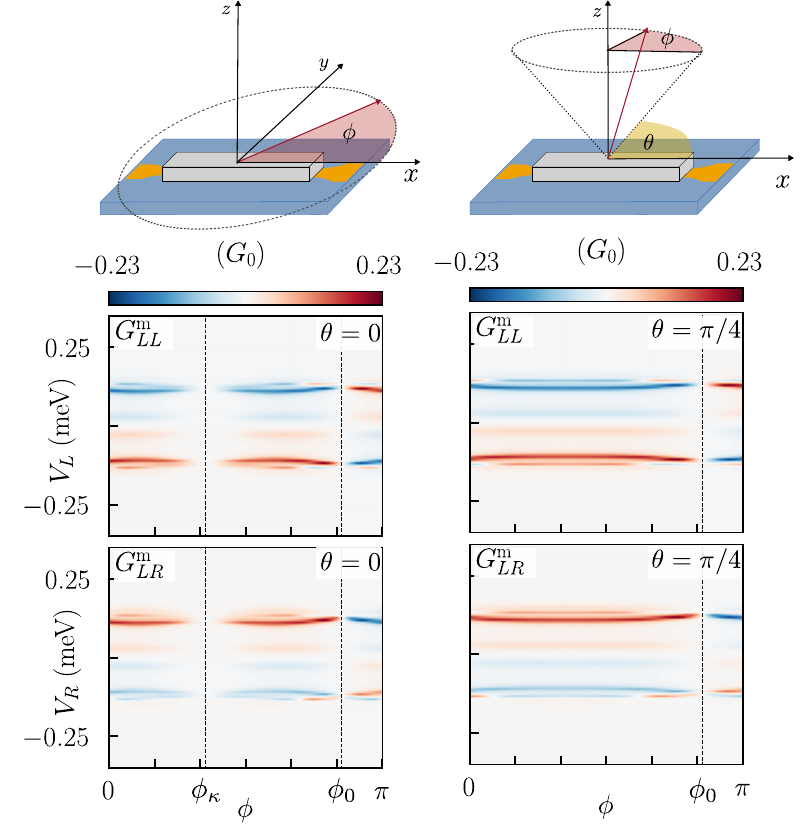}
    \caption{
    Identification of the spin-orbit orientation.
    We show how local ($G^\mathrm{m}_{LL}$) and nonlocal ($G^\mathrm{m}_{LR}$) conductance magnetic asymmetry,  Eq.~\eqref{eq:G^m_ab}, can be used to identify the spin-orbit coupling orientation by measuring these quantities while rotating the magnetic field, i.e. the angles $\qty(\theta, \phi)$ in the reference frame introduced in  Fig.~\ref{fig:supersem_soc_sketches}. 
    Both these quantities vanish linearly when the condition $\vb{b}_\bot \cross \vb*{\kappa} = 0$ is met. In the case shown, the Rashba spin-orbit coupling $\alpha_R$ is oriented in the $z$-direction and the relative strength of Dresselhaus over Rashba spin-orbit coupling is $\beta/\alpha_\perp=1/2= \tan(\phi_\kappa)$. Therefore $G^\mathrm{m}$ vanishes only for the plane identified by the angle $\phi_0 = \phi_\kappa + \pi/2$ and is thus observed both in the case of $\theta=0$ shown in the left panels and the case $\theta=\pi/4$ shown in the right ones. 
   In this simulation, $V_{\mathrm{b},L}=\SI{0.3}{\milli\eV}$, $V_{\mathrm{b},R}=\SI{0}{\micro\eV}$, $\beta=\SI{5}{\milli\eV \nano\meter}$, while the Zeeman field strength is $b=\SI{0.42}{\milli\eV}$.}
    \label{fig:magnetic_phi}
\end{figure}

The reciprocal conductances $G^\mathrm{ra}$ and $G^\mathrm{re}$ can be used to characterize the spin-orbit coupling of the nanowire. Indeed both are symmetric under reversal of magnetic field only if the system satisfies an antiunitary symmetry that persists at a finite magnetic field. An alternative and easier measurement is the conductance magnetic asymmetry $G^\mathrm{m}$, Eq. \eqref{eq:G^m_ab}, since it requires the combination of only two differential conductances at the same terminal. This quantity vanishes if there is an antiunitary symmetry persisting at a finite magnetic field. Measuring this quantity while rotating the azimuth of the magnetic field allows for the identification of the direction of the generalized spin-orbit coupling vector $\vb{e}_\kappa$, as shown in Fig.~\ref{fig:magnetic_phi}. 

The zero of the quantity $G^\mathrm{m}(\phi = \phi_0) = 0$ is achieved only when the orthogonality condition $\vb{b} \cdot \vb*{\kappa} = 0$ is satisfied. Therefore, with the measured set of directions for which $G^\mathrm{m}=0$ it is possible to determine $\vb{e}_{\kappa}$ and its relative angle with the wire direction $\phi_\kappa = \phi_0 - \pi/2$. With this information, it is possible to determine both the direction of the Rashba field and the ratio of the orthogonal Rashba and Dresselhaus SOC. The orthogonal Rashba field $\vb*{\alpha}_\perp$ is oriented in the direction $\vb{e}_\kappa\cross \vb{e}_{x}$ while the ratio of the two fields is connected to the angle by $\beta/\alpha_\perp = \tan(\phi_\kappa)$.

Note that $G^\mathrm{m}(\phi)$ shows a linear behavior in $\phi$ near $\phi_0$ (marked by a change of sign in the neighborhood). In the simulations we noticed an additional zero in the direction $(\phi_\kappa, \theta_\kappa)$ that is  $\vb*{\kappa} \cross \vb{b} = 0$. In this case, $G^\mathrm{m}(\phi, \theta)$ has a quadratic behavior in both $\phi$ and $\theta$ in the neighborhood of $(\phi_\kappa,\theta_\kappa)$. Note that $\theta_\kappa=0$ in the chosen coordinate system. 

\subsubsection{Finite-bias effect and dissipation}
\label{sec:num_wire_nonideal}

\begin{figure}[!ht]
    \centering
    \includegraphics[width=\columnwidth]{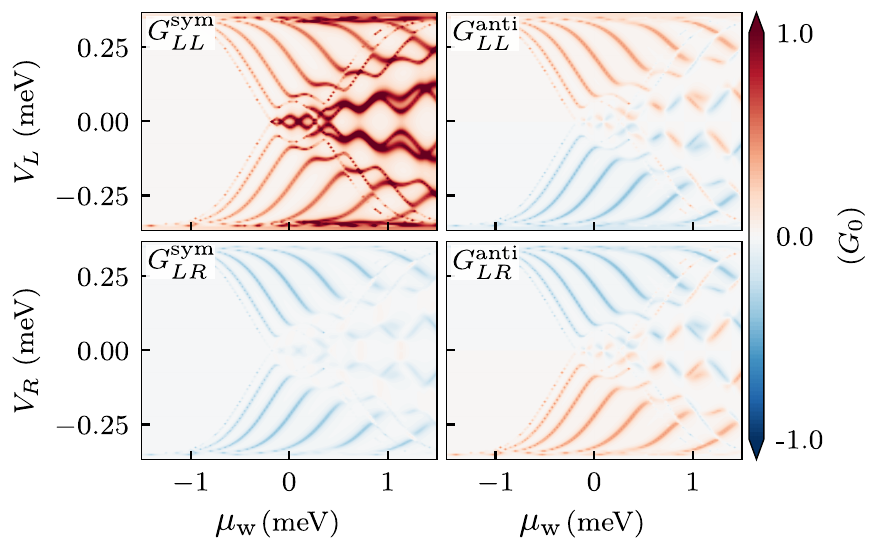}
    \caption{Splitting into symmetric and antisymmetric parts of electric differential conductance. The antisymmetric parts of local and nonlocal electric conductance are opposite. This corresponds to the symmetry relation $G^\mathrm{sa}(V) \equiv G^\mathrm{anti}_{LL}(V)+G^\mathrm{anti}_{LR}(V)=0$ in ideal systems. Deviations from this symmetry relation can be used to identify non-ideal effects like dissipation or finite-bias effect. 
    }
    \label{fig:sym_and_anti}
\end{figure}

Transport symmetries can be also exploited to assess the presence of non-ideal effects and possibly distinguish between them. To illustrate the idea in this example system, we consider the dissipation and finite-bias effect. Indeed, in an ideal system the antisymmetric components of the local and nonlocal electrical differential conductance as a function of bias voltage are opposite to each other, such that $G_\alpha^\mathrm{sa}(V) \equiv G^\mathrm{anti}_{LL}(V)+G^\mathrm{anti}_{LR}(V) = 0$ [c.f. Eq.~\eqref{eq:G_a^sa}]. This is illustrated in Fig.~\ref{fig:sym_and_anti}.

This symmetry relation is broken by finite-bias effects, dissipation, and Coulomb scattering between quasiparticles. We verify the possibility of distinguishing between finite-bias and dissipation effects by calculating the quantity $G_L^\mathrm{sa}$ in presence of these effects. We consider the same system as before in Sec. \ref{sec:num_wire_SOC} with the only difference of considering symmetric barrier of $\Delta V_\mathrm{b} = \SI{30}{\micro\eV}$ and we set the Dresselhaus SOC $\beta = 0$.  

\begin{figure}[!ht]
    \centering
    \includegraphics[width=\columnwidth]{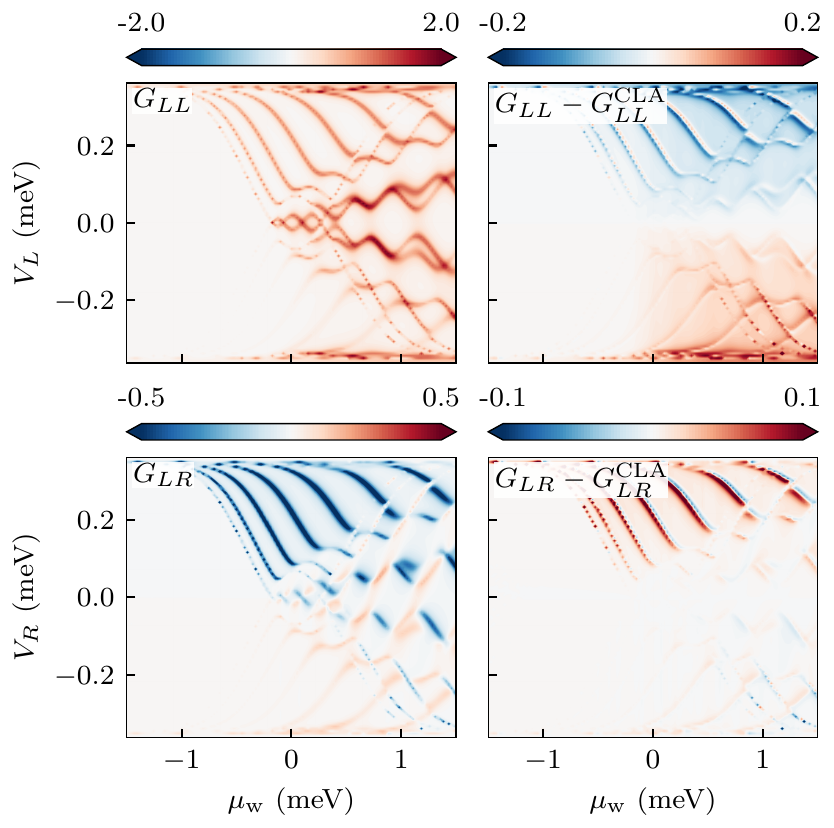}
    \caption{Electric differential conductance in the symmetric setup calculated with nonlinear theory (left panels) and correction to differential conductance in CLA. It is possible to distinguish two contributions. One general background correction in the local conductance that can be attributed to an increase in the average barrier height as the potential is raised. On top of this, we can identify a correction that moves and changes the position of the peaks. The finite-bias effect gets stronger and more evident as the barrier length is increased. In this simulation, $V_{\mathrm{b},L}=\SI{30}{\micro\eV}$, $V_{\mathrm{b},R}=\SI{30}{\micro\eV}$, $\beta=0$, while we set the Zeeman field to $\vb{b} = (1, 0, 0) \SI{40}{\micro\eV}$. 
    }
    \label{fig:nonlinear}
\end{figure}

First, we introduce the finite-bias effect by manually introducing the voltage drop in the barrier regions as shown in Fig.~\ref{fig:nanowire_sketch}. The nonlinear differential conductance is then obtained by numerical differentiation of the total current calculated with Eq.~\eqref{eq:nlqcurrent}. The comparison of the full-nonlinear theory and the CLA can be seen in Fig.~\ref{fig:nonlinear}. It is possible to distinguish two corrections, one general background correction in the local conductance that can be attributed to an increase in the average barrier height as the potential is raised. On top of this, we can identify a shift in the position of the peaks. The effect of a finite bias gets stronger and more evident as the barrier length is increased since the effect of the voltage drop is distributed in a greater area of the device.

To introduce dissipation, we compare two cases: in the first case we consider an additional normal lead acting as a quasiparticle reservoir described by a self-energy $\Sigma_\mathrm{Sm}$ of the form Eq. \eqref{eq:numerics_self_energy_n} with parameter $\gamma_n$ describing the coupling strength between system and reservoir, while in the second case we introduce a soft gap in the superconductor through the parameter $\Gamma_\nu = \Gamma_\mathrm{Sc}$ in the Dynes model for the self-energy $\Sigma_\mathrm{Sc}$ of proximity induced superconductivity, Eq. \eqref{eq:numerics_self_energy}. Note that the first case has been already assessed in Ref.~\cite{Liu_PRB_2017} for local conductance.

We consider two leads such that $\Sigma = \Sigma_\mathrm{Sc} + \Sigma_\mathrm{Sm}$. In the quasiparticle reservoir case we assume $\Gamma_\mathrm{Sc} = 0$ and $\gamma_\mathrm{Sm} = \SI{5}{\micro\eV}$ while in the soft gap case $\Gamma_\mathrm{Sc} = \SI{5}{\micro\eV}$ while $\gamma_\mathrm{Sm} = 0$. 

\begin{figure}[!hb]
    \centering
    \includegraphics[width=\columnwidth]{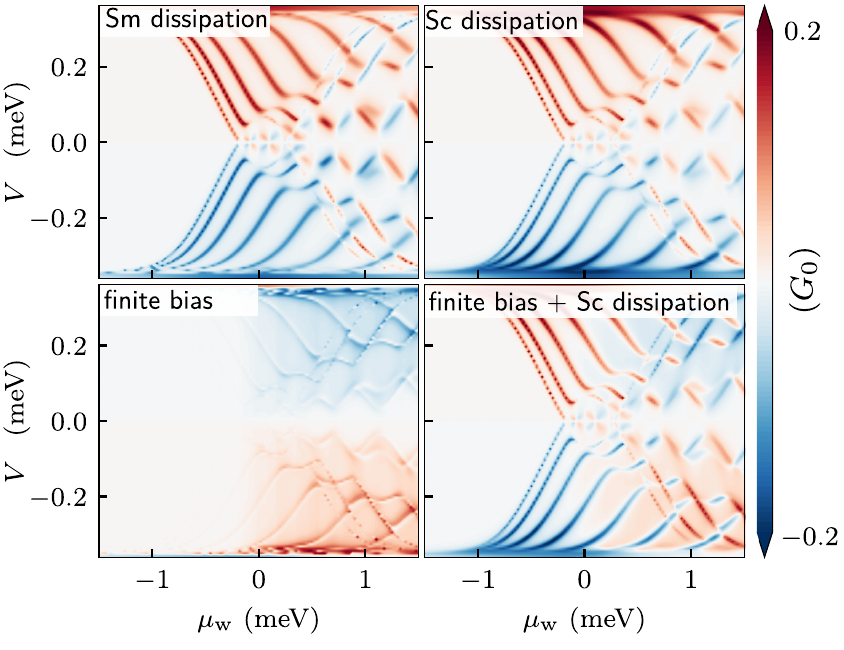}
    \caption{The plots show the value of $G^\mathrm{sa}_L$ when different symmetry-breaking mechanisms are introduced. In the upper plots, two distinct dissipation terms are considered within CLA, in the left $\gamma_\mathrm{Sm} = \SI{5}{\micro\eV}$ while in the right one $\Gamma_\mathrm{Sc} = \SI{5}{\micro\eV}$. The lower plots shows the same quantity calculated using the nonlinear theory on the left $\gamma_\mathrm{Sm} = 0$ while on the right $\Gamma_\mathrm{Sc} = \SI{5}{\micro\eV}$.
    }
    \label{fig:gsa_nonideal}
\end{figure}

The left and right plots in the top row of Fig.~\ref{fig:gsa_nonideal} shows, within CLA, the effect of a dissipation term in the wire Hamiltonian (setting $\gamma_\mathrm{Sm}=\SI{5}{\micro\eV}$, $\Gamma_\mathrm{Sc} = 0$) and of inelastic scattering processes in the superconductor modeled with the Dynes model (setting $\gamma_\mathrm{Sm} = 0$, $\Gamma_\mathrm{Sc}=\SI{5}{\micro\eV}$), respectively. The effects of the two dissipation terms are very similar and consistent with the result in Eq.~\eqref{eq:gsa_diss}. Therefore, it is not possible to distinguish between the two effects with this measurement. In the lower plots, nonlinear theory within perfect metallic screening approximation is considered. The effect of finite bias on the symmetry relation appears qualitatively different from dissipation also in this case. It can be described by a background contribution that depends on the sign of the applied voltage together with a localized correction in the position of the peaks. More strikingly, after the topological transition, there is no evident oscillation in the sign connected to the local BCS charge like in the dissipation case. Therefore measurements of electrical conductance offer the possibility of distinguishing between the effect of finite bias and dissipation. These results are consistent with previous analyses~\cite{Liu_PRB_2017, Melo_SciPost_2021}.

\subsubsection{Thermoelectric conductance}

The clear advantage of thermoelectric conductance is that temperature-induced charge accumulation, which leads to potential change modification, can be safely ignored in the regime of interest. Therefore it represents an alternative measurement free of problems related to the finite-bias effect. We stress here that by thermoelectric measurements we mean the measurement of the current as a change in the temperature of the leads. We note that for negligible inelastic scattering we expect no local thermalization, such that the device parameters should remain unchanged by the temperatures in the leads.

\begin{figure}[!hb]
    \centering
    \includegraphics[width=\columnwidth]{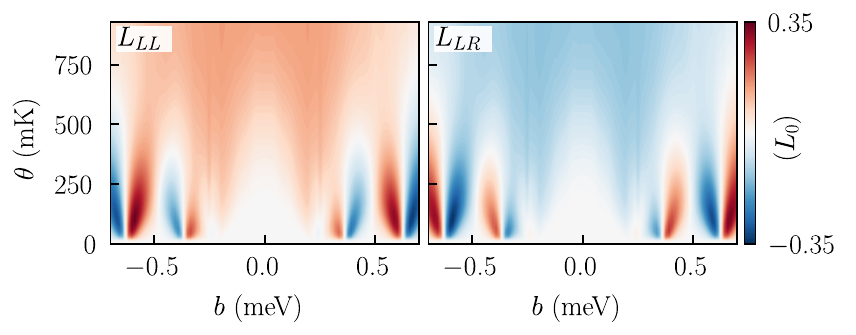}
    \caption{Local and nonlocal thermoelectric conductance in a proximitized semiconductor nanowire. The low-temperature lobes with alternating signs can be associated with the BCS charge $\expval{\tau_z}$ similarly to the interpretation of $G^\mathrm{anti}$.
    }
    \label{fig:thermoelectric}
\end{figure}

As in the case of electric differential conductance, we can define local and nonlocal thermoelectric conductance. These satisfy the symmetry relation
\begin{equation}
    L^\mathrm{sum}_L(\theta) = L_{LL}(\theta_L = \theta) + L_{LR}(\theta_R = \theta) \simeq 0 
\end{equation}
if we restrict the integral over the energy to values below the parent gap region. The latter just introduces a non-exactly balanced background contribution. As can be seen in Fig. \ref{fig:thermoelectric}, the interesting features are the lobes with an oscillating sign at low temperatures. These features can be linked to the BCS charge $\expval{\tau_z}$ of the Andreev bound states at the end of the wire by a straightforward extension of the derivation using non-local electric conductance presented in Ref. \cite{Danon_PRL_2020}.

\begin{figure}[!hb]
    \centering
    \includegraphics[width=\columnwidth]{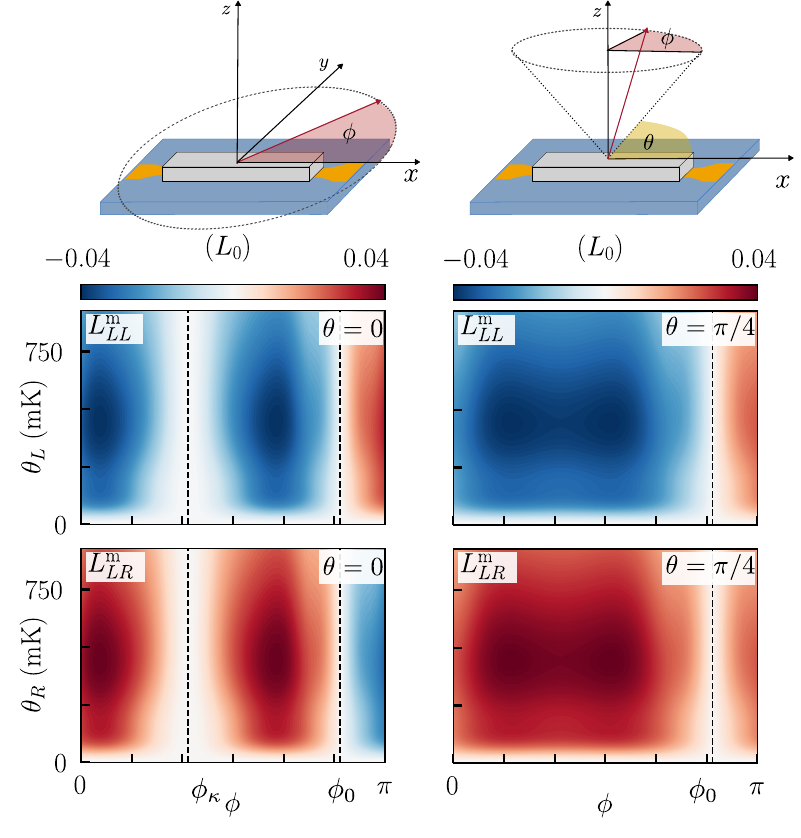}
    \caption{
    Identification of the spin-orbit orientation by thermoelectric measurements. We show how local ($L^\mathrm{m}_{LL}$) and nonlocal ($L^\mathrm{m}_{LR}$) thermoelectric conductance magnetic asymmetry can be used to identify the spin-orbit coupling orientation by measuring these quantities while rotating the magnetic field, i.e. the angles $\qty(\theta, \phi)$ in the reference frame introduced in  Fig.~\ref{fig:supersem_soc_sketches}. 
    Both these quantities vanish when the condition $\vb{b}_\bot \cross \vb*{\kappa} = 0$ is met. In the case shown, the Rashba spin-orbit coupling $\alpha_R$ is oriented in the $z$-direction while $\beta/\alpha_\perp= \tan(\phi_\kappa)=1/2$. Therefore $L^\mathrm{m}$ vanishes only for the plane identified by the angle $\phi_0 = \phi_\kappa + \pi/2$ and is thus observed both in the case of $\theta=0$ shown in the left panels and the case $\theta=\pi/4$ shown in the right ones. The parameters are the same as Fig.~\ref{fig:magnetic_phi}.
    }
    \label{fig:L_m_phi}
\end{figure}

Finite-bias effects can also affect the procedure for the determination of the spin-orbit coupling outlined in Section \ref{sec:num_wire_SOC}. The same information can be obtained by thermoelectric measurements by evaluating the thermoelectric conductance magnetic asymmetry $L^\mathrm{m}$ while rotating the magnetic field as shown in Fig.~\ref{fig:L_m_phi}. As expected, when the magnetic field lies in the plane orthogonal to the generalized spin-orbit coupling vector $\vb*{\kappa}$, identified by the angle $\phi_0$, we observe a zero in $L^\mathrm{m}$. In contrast to the electric conductance combination, $G^m_{\alpha \beta}$, the thermoelectric conductance combination $L_{LR}^m$ displays a quadratic behavior in $\phi$ around $\phi_0$ at the magnetic field angle $\theta=\pi/4$. Also for $L^m$ we observe an additional quadratic zero at $(\phi_\kappa, \theta_\kappa)$, i.e. when $\vb*{\kappa} \cross \vb{b} = 0$.

\section{Conclusions}
In this work, we have explored the limits of local and nonlocal tunneling spectroscopy of hybrid devices within the extended Landauer-Büttiker formalism. We have derived symmetry constraints on the multiterminal conductance matrix that follow from the fundamental microreversibility and particle-hole conjugation in the presence of superconductivity. Our first result shows that the reciprocal conductances $G^\mathrm{ra}$ and $G^\mathrm{re}$, defined in Eqs.~\eqref{eq:G_ab^ra} and \eqref{eq:G_ab^re}, can be employed to extract the antisymmetric-in-voltage parts of the individual electron and Andreev transmission and reflection probabilities. 

In the presence of an additional antiunitary symmetry that persists at a finite Zeeman field, a further relation can be derived for the conductance magnetic asymmetry $G^\mathrm{m}$ in Eq.~\eqref{eq:G^m_ab}. This relation is particularly useful in the study of spin-orbit coupled semiconductor nanowires proximitized by an $s$-wave superconductor since it allows extracting the ratio between the Rashba and Dresselhaus spin-orbit coupling strength. 
We have demonstrated this result in an explicit numerical model. This result may be useful for material and device characterization because the characterization of the spin-orbit coupling in proximitized semiconductor devices is an open research question. Future work can study these quantities in a more realistic scenario modeling the cross-section of the superconductor-semiconductor heterostructure to include multiple transverse modes and the orbital coupling of the magnetic field.

Furthermore, we have studied the effects of dissipation and the dependence of the electric potential on the bias voltage on the symmetry relations at an explicit model of a proximitized semiconductor nanowire. Generally, these symmetry relations are broken by these non-idealities. However, the two effects yield distinct signatures in the conductance matrix elements and their linear combinations. 

In conclusion, nonlocal tunneling spectroscopy is a powerful tool employed in the study of Andreev bound states \cite{Puglia_PRB_2021, Menard_PRL_2020,Danon_PRL_2020, Sbierski_PRB_2022,Banerjee_arXiv_2022a, Banerjee_arXiv_2022b, Banerjee_arXiv_2022c}. We hope that our work contributes to the interpretation of the experimental measurements and expands the scope of the method by allowing access to more detailed information on the system under study.

\section{Acknowledgments}
The authors want to thank Alisa Danilenko, Andreas Pöschl, and Charles Marcus for useful discussions. This work was supported by the Danish National Research Foundation, the Danish Council for Independent Research \textbar Natural Sciences. The authors acknowledge Microsoft research for support and computational resources.

\bibliography{bibliography.bib}
\end{document}